\documentclass[
 reprint,
 amsmath,amssymb,
 aps,
]{revtex4-2}

\usepackage{graphicx}
\usepackage{dcolumn}
\usepackage{bm}
\usepackage{hyperref}
\usepackage{booktabs,tabularx}
\usepackage{multirow}
\usepackage{booktabs, threeparttable}
\usepackage{orcidlink}
\usepackage{array}
\usepackage{siunitx}
\usepackage{xparse}

\newcolumntype{L}[1]{>{\raggedright\arraybackslash}p{#1}}

\hypersetup{
    colorlinks=true,
    linkcolor=magenta,
    filecolor=magenta,      
    urlcolor=blue,
    citecolor=blue,
    }

\newcommand{\myItemHandler}[1]{%
  \IfStrEq{#1}{}{}{\ignorespaces#1\unskip}%
}

\begin{document}


\title{Probing non-perturbative QCD aspects on particle production in pp collisions with forward-backward correlations using \texttt{PYTHIA8}}

\author{Rohit Agarwala \orcidlink{0009-0009-8413-1123}}
 \email{therohitagarwala@buniv.edu.in}
\author{Kalyan Dey \orcidlink{0000-0002-4633-2946}}
 \email{kalyan.dey@buniv.edu.in (corresponding author)}
\affiliation{Department of Physics, Bodoland University, Kokrajhar - 783370, B.T.R Assam, India}

\date{\today}

\begin{abstract}

We used \texttt{PYTHIA8} simulations to study forward-backward (FB) correlations in proton-proton (pp) collisions at LHC energies, focusing on non‑perturbative QCD dynamics, specifically color‑reconnection (CR) mechanisms and QCD radiations (ISR/FSR).
Using \texttt{PYTHIA8} (v8.311) under ALICE/ATLAS kinematics, we analyze \textit{extensive}: FB multiplicity correlation strength, $b_{\rm corr}(mult.)$ and FB summed transverse momentum correlation strength, $b_{\rm corr}^{\sum p_{\rm T}}$; \textit{intensive}: FB mean transverse momentum correlation strength, $b_{\rm corr}^{\overline p_{\rm T}}$, and \textit{strongly intensive} ($\Sigma_{\rm N_{\rm F} \rm N_{\rm B}}$) observables in symmetric pseudorapidity intervals and validate the results against available experimental data. $b_{\rm corr}(mult.)$, when systematically evaluated as a function of the pseudorapidity gap ($\eta_{\rm gap}$ and $\eta_{\rm sep}$), as well as in different azimuthal sectors ($\eta-\varphi$), revealed that multi-parton interaction (MPI)-based CR  with CR ranges 3.6 and 5.4 and QCD Color Rope model significantly improved agreement with data. Initial- and final-state radiation (ISR, FSR) are also found to critically influence $b_{\rm corr}(mult.)$ and $b_{\rm corr}^{\sum p_{\rm T}}$, with ISR exerting a stronger effect. Disabling ISR and/or FSR fails to replicate the $\eta_{\rm gap}$-dependent trend of $b_{\rm corr}(mult.)$ and $b_{\rm corr}^{\sum p_{\rm T}}$, emphasizing their necessity.  $b_{\rm corr}^{\sum p_{\rm T}}$ shows minimal sensitivity to CR range but surges in magnitude when CR is disabled. Intensive quantities, such as the FB mean transverse momentum correlation, $b_{\rm corr}^{\overline p_{\rm T}}$, exhibit the opposite behavior to extensive observables i.e. the magnitude of $b_{\rm corr}^{\overline p_{\rm T}}$ increases with the increase of CR strength, highlighting the distinct influence of CR on intensive versus extensive observables.
Further, our analysis of the azimuthal dependence of $b_{\rm corr}^{\overline p_{\rm T}}$ provides compelling evidence that parton showers dominate short-range correlation mechanisms.
Last but not the least, for strongly intensive observables, \texttt{PYTHIA8} qualitatively reproduces the energy-dependent rise of $\Sigma_{\rm N_{\rm F} \rm N_{\rm B}}$ with $\eta_{\rm sep}$, consistent with preliminary ALICE results, though quantitative discrepancies persist. Among the QCD radiations, FSR is seen to stronger impact on this variable.
\end{abstract}

\maketitle

\section{Introduction} \label{section1}

Colliding heavy ions at ultra-relativistic energies provides an opportunity to explore the properties of strongly interacting matter under extreme conditions in the laboratory~\cite{shuryak1980,mclarren1986,bbback2005phobos,adams2005star,adcox2005phenix}. Correlations among produced particles in configuration and momentum space across diverse collision systems - from pp to nucleus-nucleus (AA) collisions - are believed to provide crucial insights into the dynamical processes governing multiparticle production. Correlations among the final-state particles can be categorized into short-range correlations (SRCs) and long-range correlations (LRCs)~\cite{ua5alner1987}. SRCs are believed to originate from various localized short-range effects, such as correlations due to minijets, decays of clusters and resonances, which are typically constrained over a smaller pseudorapidity range,  ($\eta_{\rm gap}\leq 1$)~\cite{uhlig1978}. LRCs, on the other hand, are extended over a much wider $\eta_{\rm gap}$ (typically $>$ 1 unit) and are believed to originate from initial-state fluctuations in the number of particle-emitting sources, such as the number of participating partons, multi-parton interactions (MPI), or cut Pomerons, which serve as effective representations of MPI in Regge-based models~\cite{werner2008,werner2014,bravina2018}. LRCs may also emerge from the collective expansion of the Quark-Gluon Plasma (QGP), leading to correlated particle emission over large rapidity intervals.
The study of forward-backward (FB) correlations provides a means to disentangle SRCs from LRCs~\cite{capella1994,capella1978,vechernin2007multiplicity,braun2004,braun2000implications}. FB correlations are typically quantified by measuring the correlations among observables such as particle multiplicity, summed transverse momentum, and average transverse momentum,  within symmetrically separated pseudorapidity intervals. Both FB multiplicity correlations and summed transverse momentum correlations are extensive observables, scaling directly with the number of particle-emitting sources, which may vary with system size, event multiplicity, or centrality. These correlations are also sensitive to volume fluctuations. In contrast, the FB mean transverse momentum correlation is an intensive observable, normalized by the particle multiplicity in each pseudorapidity region, and is therefore independent of the system size or total multiplicity. This property makes it a robust tool to probe intrinsic particle production dynamics, free from volume fluctuation effects. In addition to intensive observables, there exists a more robust class of observables known as \emph{strongly intensive variables}. Among these, the quantity $\Sigma$ is particularly notable, as it exhibits no dependence on the system volume or its fluctuations~\cite{gorenstein2011,anticic2015}. \\

Early measurements of FB multiplicity correlations in symmetric pseudorapidity intervals were first performed at the CERN ISR for pp collisions, reporting a positive correlation strength~\cite{albrow1978}. Similar findings were later observed in $p \bar p$ collisions at the CERN SPS across $\sqrt{s}$ = 200–900 GeV~\cite{alpgard1983, ua5alner1987, ansorge1988}. The FB correlation in the heavy-ion sector was measured for the first time by the STAR experiment at RHIC, revealing strong long-range correlations (LRCs) in Au+Au collisions~\cite{abelev2009,bksrivastava2007,srivastava2008star2,tarnowsky2007}. Later measurements by ATLAS~\cite{aad2012atlas} and ALICE~\cite{adam2015} at the LHC confirmed strong FB correlations in pp collisions, with correlation strength increasing as a function of both energy and system size. FB multiplicity and mean transverse momentum correlation were studied for the first time at LHC energies with Pb+Pb collisions at $\sqrt{s_{\rm NN}}$ = 2.76 TeV and 5.02 TeV~\cite{altsybeev2017}. While the FB multiplicity correlation strength depends on both the centrality estimator and the width of the centrality class, the FB mean transverse momentum correlation remains unaffected by these factors, highlighting its robustness as a probe of collision dynamics. These observations have since motivated extensive theoretical and phenomenological investigations~\cite{armesto2007,konchakovski2009,brogueira2009,bzdak2009,yan2009,yan2010,lappi2010,bialas2011,mondal2020,mondal2023}. \\

Regge theory-based models, including the Dual Parton Model (\texttt{DPM})~\cite{capella1994} and Quark-Gluon String Model (\texttt{QGSM})~\cite{kaidalov2003}, attribute long-range correlations (LRCs) observed in pp collisions to multi-string dynamics governed by fluctuations in cut Pomerons~\cite{capella1978,bravina2018}. Extensions such as the String Fusion Model (\texttt{SFM}) and string percolation framework incorporate interactions among overlapping strings to improve LRC descriptions in high-multiplicity pp collisions~\cite{amelin1993plb,amelin1994prl,amelin1994zpc,brogueira2007}. Subsequent studies in pp collisions~\cite{khan2019} with models like \texttt{EPOS1.99}, \texttt{QGSJETII-04}, and hybrid approaches (e.g., \texttt{EPOS3} without hydrodynamic evolution)~\cite{mondal2020} also failed to reproduce FB multiplicity and summed transverse momentum correlation data. The \texttt{UrQMD}, a microscopic transport model, has also been employed to study FB correlations in pp collisions~\cite{bhattacharyya2024}. Although it succeeds in reproducing experimental trends, the quantitative description of the data was not met. However, the FB correlation \texttt{UrQMD} results were found to be in qualitative agreement with \texttt{QGSM} and \texttt{EPOS3}, studied by some other group of researchers~\cite{bravina2018,mondal2020}. \\

Studies of string-based models, including \texttt{PYTHIA6} (MC09, DW, AMBT2B, Perugia tunes), \texttt{PYTHIA8} 4C, and \texttt{HERWIG++}, reveal broad discrepancies with ATLAS forward-backward (FB) correlation data for pp collisions, barring \texttt{PYTHIA6} AMBT2B, which exhibits limited consistency at smaller $\eta$ ranges~\cite{aad2012atlas}. Moreover, models like \texttt{PHOJET} and Perugia-tuned \texttt{PYTHIA6} continue to exhibit discrepancies with ALICE data on pp collisions~\cite{adam2015}, underscoring limitations in the traditional parton shower and hadronization approaches. \\

Recent investigations using \texttt{PYTHIA8} showed that color reconnection (CR) affects the FB multiplicity correlation in pp collisions at LHC energies; however, default configurations remain insufficient
~\cite{kundu2019,cuautle2019}. Although tuned CR parameters along with specific MPI ranges lead to modest improvements, quantitative agreement with the experimental data remains unattained. This motivates us to further explore and tune CR and other relevant parameters in \texttt{PYTHIA} to achieve a better quantitative description of the observed FB correlations. A similar line of investigation has been carried out using different tunes of \texttt{PYTHIA}: this time focusing on the description of the strongly intensive observable $\Sigma$. Calculations performed with the \texttt{PYTHIA8} Monash 2013 Tune (w/o CR) and \texttt{EPOS3} model in pp collisions~\cite{erokhin2021} ($\sqrt{s}$ = 0.9, 2.76, 5.02, 7 and 13 TeV) reveal systematic deviations from preliminary ALICE data, further underscoring the limitations of existing model configurations in quantitatively reproducing the measured correlation patterns. In contrast, models incorporating explicit string fusion and percolation mechanisms, such as a phenomenological string model~\cite{andronov2019} demonstrate partial agreement, particularly at pseudorapidity separations greater than unity. Furthermore, the \texttt{EPOS1.99} model has been shown to successfully describe the NA61/SHINE data on $\Sigma[N_{\rm f},N_{\rm b}]$ in pp collisions at lower SPS energies ($\sqrt{s}\sim$ 6-17 GeV) ~\cite{prokhorova2018pseudorapidity}. Similarly, the hybrid hydrodynamics-based Monte Carlo generator \texttt{HYDJET++} has been applied to Pb+Pb collisions to study FB multiplicity and FB mean transverse momentum correlations~\cite{malik2024,singh2024}, but the simulations exhibit no qualitative agreement with ALICE measurements~\cite{altsybeev2017}. \\

Further, the strongly intensive observable $\Sigma$ has also been investigated in heavy-ion collisions (Pb+Pb at $\sqrt{s_{\rm NN}}$ = 2.76 and 5.02 TeV and Xe+Xe at $\sqrt{s_{\rm NN}}$ = 5.44 TeV) within the wounded nucleon (WNM) and wounded quark (WQM) models~\cite{sputowska2023forward}. While conventional models such as \texttt{HIJING}, \texttt{AMPT}, and \texttt{EPOS} fail to describe ALICE measurements~\cite{sputowska2019forward,sputowska2022forward}, both WNM (qualitatively) and WQM (quantitatively) describe the ALICE results. The success of WNM and WQM lies in the fact that these frameworks model particle production through wounded constituents, incorporate fluctuations in their number, and directly link $\Sigma$ to the fragmentation function via the probability parameter (p), which tells that a wounded nucleon emits a particle in a given pseudorapidity interval~\cite{sputowska2023forward}. These features enable them to correctly reproduce the centrality and energy dependence of $\Sigma$ as observed in ALICE measurements. In the context of pp collisions, as discussed above, \texttt{PYTHIA8} (w/ \& w/o CR) has not been able to interpret Preliminary ALICE results. Henceforth, in this present investigation, an attempt is made to estimate this quantity ($\Sigma$) for different variants of the \texttt{PYTHIA8} framework, like altering CR ranges, enabling/disabling parton showers, and taking into account various hadronization mechanisms in terms of CR and QCD Color Ropes. \\

This paper is structured as follows: Section~\ref{section2} provides a brief overview of the \texttt{PYTHIA8} event generator. Section~\ref{section3} details the observables under study: the multiplicity correlation measure $b_{\rm corr}(mult.)$, transverse momentum correlations, $b_{\rm corr}^{\sum p_{\rm T}}$ and $b_{\rm corr}^{\overline p_{\rm T}}$, and the strongly intensive quantity, $\Sigma_{\rm N_{\rm F} \rm N_{\rm B}}$. Section~\ref{section4} outlines the analysis methodology and event selection criteria. Section~\ref{section5} presents the results obtained using \texttt{PYTHIA8} (version 8.311) and its various configurations, followed by Section~\ref{section6}, which summarizes the key findings.


\section{The \texttt{PYTHIA8} model} \label{section2}

\texttt{PYTHIA8} is a general-purpose Monte Carlo (MC) event generator that can simulate a variety of collision systems, like $pp$, $e^+ e^-$, etc. in three stages: the primary hard-scattering process, the subsequent parton shower, and the fragmentation of colored particles into colorless hadrons~\cite{bierlich2022}. It handles both perturbative and non-perturbative Quantum Chromodynamics (QCD) processes. Perturbative QCD (pQCD) processes govern the high-energy interactions between particles where the strong coupling constant ($\alpha_{\rm s}$) is small, including processes like jet production, top-quark pair production, Initial State Radiation (ISR), and Final State Radiation (FSR). In \texttt{PYTHIA8}, ISR and FSR are modeled as interleaved, $p_{\rm T}$-ordered showers that account for the QCD evolution of partons before and after the hard scattering. Their coherent integration with MPI ensures a consistent description of both perturbative and non-perturbative QCD dynamics, critical for simulating observables like FB correlations, and collective phenomena in high-energy collisions~\cite{skands2014,sjostrand2015}. Non-perturbative QCD (npQCD) encompasses phenomena where $\alpha_{\rm s}$ is large and necessitates phenomenological models. Key components include MPI, color reconnection (CR), hadronization via fragmentation route (Lund fragmentation), etc. \texttt{PYTHIA8} incorporates MPI to account for multiple independent partonic interactions within a single collision. MPIs are vital for reproducing observed charged-particle multiplicities and transverse momentum spectra in high-energy collisions. The CR mechanism, on the other hand, addresses the rearrangement of color strings between final-state partons to minimize potential energy. For the present study, two different CR models, namely MPI-based CR and QCD-based CR, are employed. While both mechanisms utilize color string length minimization as their foundation, they adopt distinct methodologies to reconfigure color flow during the hadronization process. Adding the SU(3) color rules from QCD during the reconnection process of strings and the insertion of junction structures is the main difference between the MPI-based and QCD-based CR models. CR mechanism plays a significant role in accounting for the non-perturbative effects that arise from the interactions between multiple parton-parton scatterings, which are a dominant feature in high-multiplicity (HM) pp collisions. Like in Ref.~\cite{kar2017}, it was concluded that ``hydrodynamic collectivity" could be explained in terms of the CR mechanism. For instance, MPI-based CR with different reconnection range (RR) could only qualitatively reproduce collective-like phenomena such as the mass-dependent hardening of transverse momentum spectra as a function of charged particle multiplicity~\cite{sarma2023}, CR could also explain $(p + \bar p)$/$(\pi^{+} + \pi^{-})$ ratio in HM pp collisions~\cite{bierlich2015,ortiz2013,dey2022}. The strength of CR is quantified by a parameter, known as the reconnection range. In this work, we elucidate the role of parton showers (ISR \& FSR) , vary the CR range parameter, and test different variants of CR models in \texttt{PYTHIA8} to probe their impact on the FB correlations and the strongly intensive observable. \\

In \texttt{PYTHIA8}, the default hadronization model is based on the Lund String Fragmentation Model~\cite{andersson1983,sjostrand1984}, where quarks and gluons are connected by strings that eventually fragment into hadrons after hadronization. But in the high-multiplicity (HM) events, multiple color strings make a cluster to constitute `color ropes' having larger string tension leading to a higher probability of producing heavier hadrons (e.g., baryons and strange particles), which is evident from the recent articles in this regard, wherein this mechanism could explain the multistrange hadron yield ratios or in other words, \textit{strangeness enhancement}~\cite{biro1984,bierlich2015strings,bianchi2016,alice2017enhanced,nayak2019,hushnud2024,acharya2024,acharya2025}. In this work, we therefore investigate the influence of the color rope hadronization mechanism on FB correlations and the strongly intensive quantity. The parameters associated with this model are provided in Tab.~\ref{tab:parameters}. A comprehensive description of the model is available in Ref.~\cite{bierlich2022}.

\begin{table}[ht]
\centering
\caption{Parameters related to QCD Color Ropes in \texttt{PYTHIA8}}
\vspace{1mm}
\begin{tabular}{l r}
\hline
\multicolumn{2}{c}{\textbf{QCD Color Ropes}} \\
\hline \hline
Parameters & Values \\
\hline
    \texttt{Ropewalk:RopeHadronization} & On \\
    \texttt{Ropewalk:doShoving} & On \\
    \texttt{Ropewalk:doFlavour} & On \\
    \texttt{Ropewalk:rCutOff} & 10 \\
    \texttt{Ropewalk:r0} & 0.5 \\
    \texttt{Ropewalk:m0} & 0.2 \\
    \texttt{Ropewalk:beta} & 0.1 \\
    \texttt{Ropewalk:tInit} & 1.5 \\
    \texttt{Ropewalk:deltat} & 0.05 \\
    \texttt{Ropewalk:tShove} & 10 \\
    \texttt{Ropewalk:gAmplitude} & 0 \\
\hline
\end{tabular}
\label{tab:parameters}
\end{table}
\begin{figure}
    \centering
    \includegraphics[width=1.\linewidth]{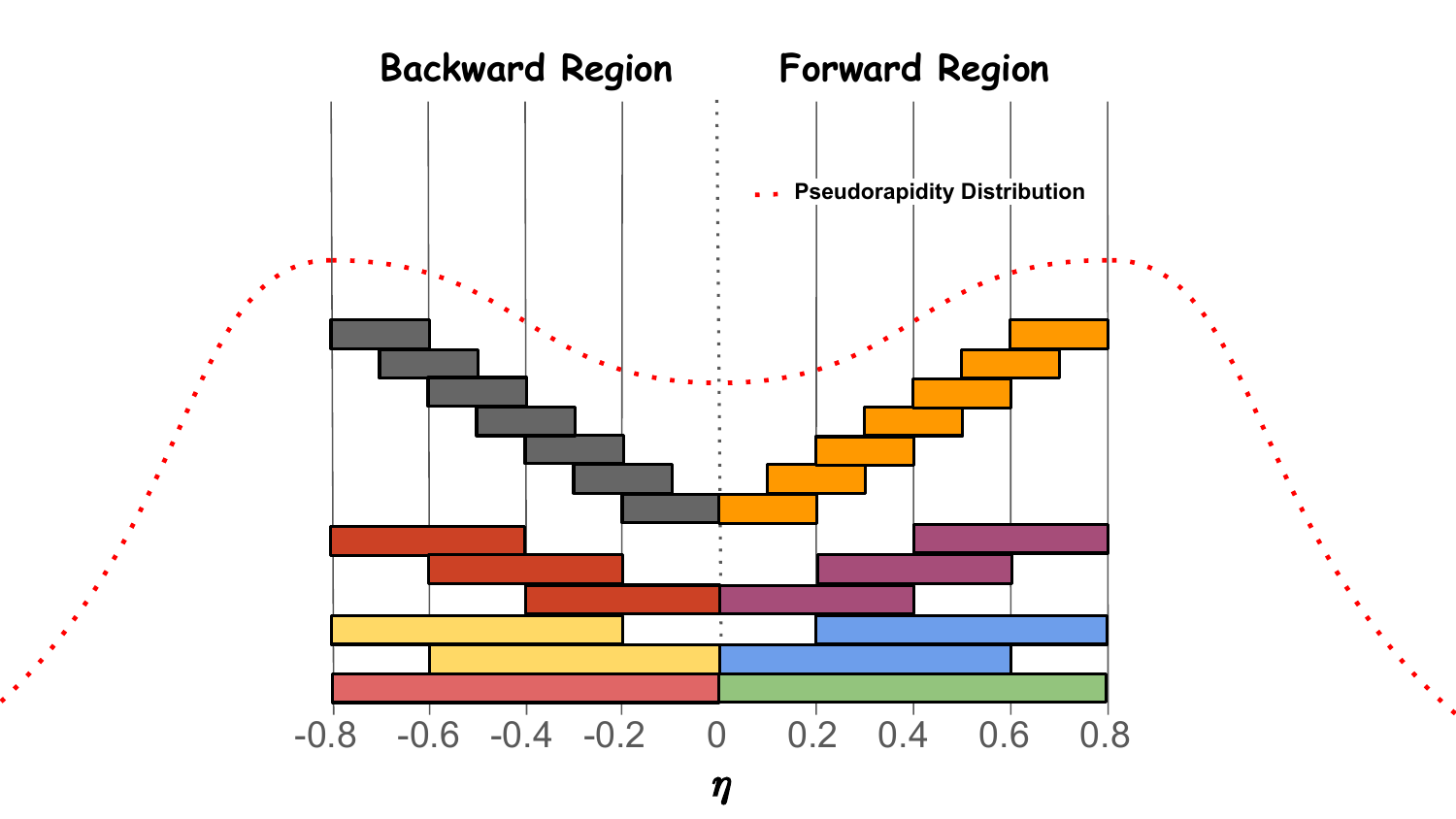}
    \includegraphics[width=1.\linewidth]{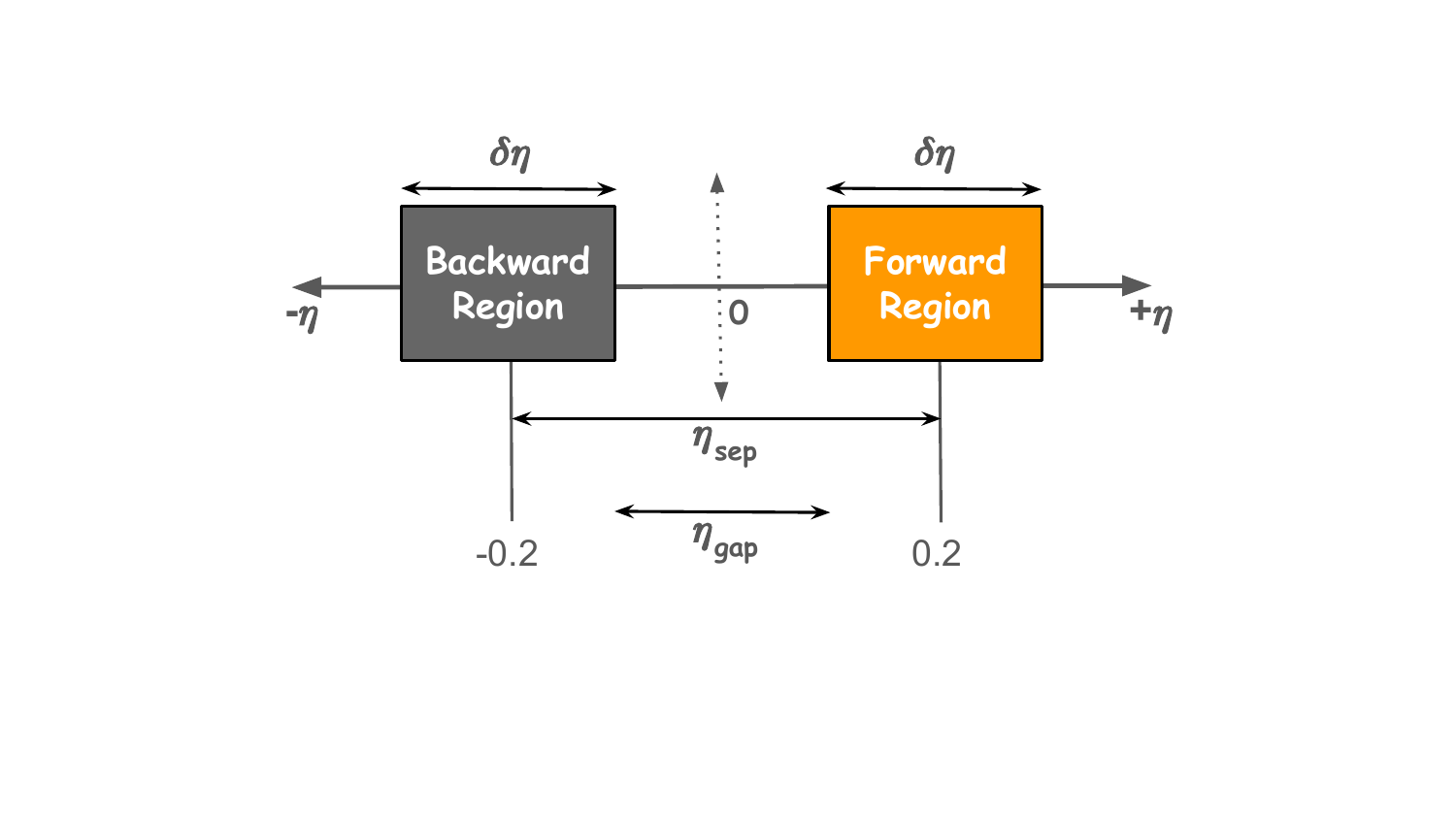}
    \vspace{-1.8cm}
    \caption{Cartoon illustrating forward (F) and backward (B) pseudorapidity distribution and windows. The figure at the bottom is a zoomed-in version of one of the F and B windows, clearly showing pseudorapidity width ($\delta\eta$), pseudorapidity gap ($\eta_{\rm gap}$), and pseudorapidity separation ($\eta_{\rm sep}$).}
    \label{fig:illustrationnew}
\end{figure}

\section{Observables of Interest} \label{section3}

\subsection{Extensive Quantities}

\subsubsection{Forward-Backward Multiplicity Correlation}

FB multiplicity correlation, denoted by $b_{\rm corr}(mult.)$, is quantified using the Pearson correlation coefficient, which is derived from a linear relationship between the event-wise average charged-particle multiplicity in the backward
$\eta$-window $\langle N_b \rangle$, and the corresponding average charged-particle multiplicity the forward $\eta$-window, $\langle N_f \rangle$~\cite{ua5alner1987},

 \begin{equation}\label{eq1}
   \langle N_{\rm b} \rangle_{N_{\rm f}} = a + b_{\rm corr}(mult.).N_{\rm f}.
   \end{equation}
The strength of the FB multiplicity correlation is therefore given by,
    \begin{equation}\label{eq2}
    b_{\rm corr}(mult.) = \frac{\left\langle N_{\rm f}N_{\rm b}\right\rangle - \left\langle N_{\rm f}\right\rangle\left\langle N_{\rm b}\right\rangle}{ \langle N_{\rm f}^2\rangle - \left\langle N_{\rm f}\right\rangle^2} 
    \end{equation}

where `$a$' represents the number of uncorrelated particles~\cite{ua5alner1987,capella1994}. One can estimate the FB multiplicity correlation strength in two ways. The first is by linear regression using Eqn.~\ref{eq1} and the other is by directly using the formula (Eqn.~\ref{eq2}). For the present estimation, the formula method is used. An illustration of pseudorapidity ($\eta$) distribution with various slices of forward and backward windows is shown in Fig.~\ref{fig:illustrationnew}. \\

Although forward-backward (FB) multiplicity correlations are useful for distinguishing short-range (SRC) and long-range (LRC) correlations, their extensive nature makes them susceptible to system size dependence and volume fluctuations. This work will therefore also examine three alternative observables (i) FB summed transverse momentum correlation ($b_{\rm corr}^{\sum p_{\rm T}}$) and (ii) FB mean transverse momentum correlation ($b_{\rm corr}^{\overline p_{\rm T}}$) and (iii) a strongly intensive quantity ($\Sigma$).


\subsubsection{Forward-Backward Summed Transverse Momentum Correlation}

Similar to FB multiplicity correlation, FB summed transverse momentum correlation is an extensive observable, that refers to the correlation between summed transverse momenta of produced particles in the forward
$\langle \sum p_{\rm T}^f \rangle$ and in backward $\langle \sum p_{\rm T}^b \rangle$ hemispheres. Similar to the definition of the FB multiplicity correlation, the FB summed transverse momentum correlation can be defined as,

 \begin{equation} \label{atlas_formula}
     b_{\rm corr}^{\sum p_{\rm T}} = \frac{\langle  (\sum p_{\rm T}^{\rm f} - \langle \sum p_{\rm T}^{\rm f} \rangle ) (\sum p_{\rm T}^{\rm b} - \langle \sum p_{\rm T}^{\rm b} \rangle) \rangle}{\sqrt{\langle  (\sum p_{\rm T}^{\rm f} - \langle \sum p_{\rm T}^{\rm f} \rangle )^{2} \rangle \langle (\sum p_{\rm T}^{\rm b} - \langle \sum p_{\rm T}^{\rm b} \rangle)^{2} \rangle}}.
 \end{equation} 
   
Here, $\sum p_{\rm T}^{\rm f}$ and $\sum p_{\rm T}^{\rm b}$ are the sums of the magnitudes of transverse momenta ($p_{\rm T}$) of produced charged particles in the forward and backward pseudorapidity regions, respectively. ATLAS collaboration at the LHC had used the following definition of summed transverse momentum correlation~\cite{aad2012atlas}.


\subsection{Intensive Quantities}

\subsubsection{Forward-Backward Mean Transverse Momentum Correlation}
Unlike the previous two observables, the FB mean transverse momentum correlation ($b_{\rm corr}^{\overline p_{\rm T}}$) is intensive in nature and hence remains unaffected by the system size, making it a valuable tool for probing the underlying dynamics of particle production. The central differentiating factor between $b_{\rm corr}^{\sum p_{\rm T}}$ and $b_{\rm corr}^{\overline p_{\rm T}}$ is that for the later, one needs to calculate the $\overline p_{\rm T}$ on event-by-event basis. Below is the formula to calculate the correlation strength,
\begin{equation}
    b_{\rm corr}^{\overline p_{\rm T}} = \frac{\langle \overline p^{\rm f}_{\rm T} \overline p^{\rm b}_{\rm T} \rangle - \langle \overline p^{\rm f}_{\rm T} \rangle \langle \overline p^{\rm b}_{\rm T} \rangle}{\langle \overline p^{\rm f^{2}}_{\rm T} \rangle - \langle \overline p^{\rm f}_{\rm T} \rangle^{2}},
\end{equation}
where the mean transverse momenta of produced particles in forward and backward regions are given by $\overline p_{\rm T}^{\rm f}$ = $ \sum_{i=1}^{N_{\rm f}} \frac{p_{\rm T}^{i}}{N_{\rm f}}$ and $\overline p_{\rm T}^{\rm b}$ = $ \sum_{i=1}^{N_{\rm b}} \frac{p_{\rm T}^{i}}{N_{\rm b}}$.

\begin{table*}[ht]
\centering
\caption{Kinematics used for various observables investigated in this article}
\label{tab:kinematics}
\renewcommand{\arraystretch}{1.5}
\begin{tabularx}{\textwidth}{>{\raggedright\arraybackslash}X>{\raggedleft\arraybackslash}X}
\toprule
\hline
\textbf{Observables of Interest} & \textbf{Kinematical Cuts} \\
\midrule
\textit{FB Multiplicity Correlation}, $b_{\rm corr}(mult.)$ & \textsc{ALICE} ($0.3 < p_{\rm T} < 1.5$ GeV/c, $|\eta| < 0.8$) \\
\textit{FB Summed Transverse Momentum Correlation}, $b_{\rm corr}^{\sum p_{\rm T}}$ & \textsc{ATLAS} ($p_{\rm T} > 0.1$ GeV/c, $|\eta| < 2.5$, $N_{\rm ch} \geq 2$) \\
\textit{FB Mean Transverse Momentum Correlation}, $b_{\rm corr}^{\overline p_{\rm T}}$ & \textsc{ALICE} ($0.2 < p_{\rm T} < 2.0$ GeV/c, $|\eta| < 0.8$) \\
\textit{Strongly Intensive Quantity}, $\Sigma_{\rm N_{\rm F} \rm N_{\rm B}}$ & \textsc{ALICE} ($0.2 < p_{\rm T} < 2.0$ GeV/c, $|\eta| < 0.8$) \\
\hline
\bottomrule
\end{tabularx}
\end{table*}

\subsubsection{Strongly Intensive Quantity}

Strongly intensive quantities, such as $\Delta$ and $\Sigma$, have emerged as critical tools for characterizing systems where observables are independent of the volume of the system and its fluctuations, thus isolating intrinsic dynamical correlations in particle production processes~\cite{gorenstein2011}. These quantities are rigorously defined to suppress contributions from volume fluctuations, which often dominate correlations in high-energy particle collisions~\cite{anticic2015}. Among these, the $\Sigma$ observable is particularly significant due to its explicit dependence on the covariance term, which encodes pairwise correlations between observables. In contrast to traditional intensive quantities, which may retain residual volume dependence in specific scenarios, $\Sigma$ is strictly \textit{strongly} intensive, ensuring its robustness for comparisons across collision systems with varying geometries or centralities~\cite{rustamov2017}. \\

The $\Sigma$ observable can be derived using two extensive observables, such as charged-particle multiplicities or transverse momenta, measured in distinct F and B kinematic regions. In this work, we focus on $\Sigma_{\rm N_{\rm F} \rm N_{\rm B}}$, constructed from the charged-particle multiplicities in forward ($N_{\rm f}$) and backward ($N_{\rm b}$) pseudorapidity intervals. This choice leverages the pseudorapidity dependence of correlations to probe long-range dynamics while minimizing short-range contributions from resonance decays or jets. The formula for $\Sigma_{\rm N_{\rm F} \rm N_{\rm B}}$ is given by,

\begin{equation} \label{eq5}
    \Sigma_{\rm N_{\rm F} \rm N_{\rm B}} = \frac{\omega_{\rm N_{\rm b}} \langle N_{\rm f} \rangle + \omega_{\rm N_{\rm f}} \langle N_{\rm b} \rangle - 2Cov(N_{\rm f}, N_{\rm b})}{ \langle N_{\rm f} \rangle
+ \langle N_{\rm b} \rangle}.
\end{equation}

Here, $Cov(N_{ f}, N_{\rm b}) \equiv \langle N_{\rm f} N_{\rm b} \rangle - \langle N_{\rm f} \rangle \langle N_{\rm b} \rangle$, quantifies the covariance between $N_{\rm f}$ and $N_{\rm b}$ (correlation term) and $\omega_{\rm N_{\rm f}}$ and $\omega_{\rm N_{\rm b}}$ are the scaled variances of multiplicity fluctuations, given by the formula,

\begin{equation}
    \omega_{\rm N} \equiv \frac{Var(N)}{\langle N \rangle} = \frac{\langle N^{2} \rangle - \langle N \rangle^2}{\langle N \rangle}.
\end{equation}

\begin{figure*}
    \centering
    \includegraphics[width=\linewidth]{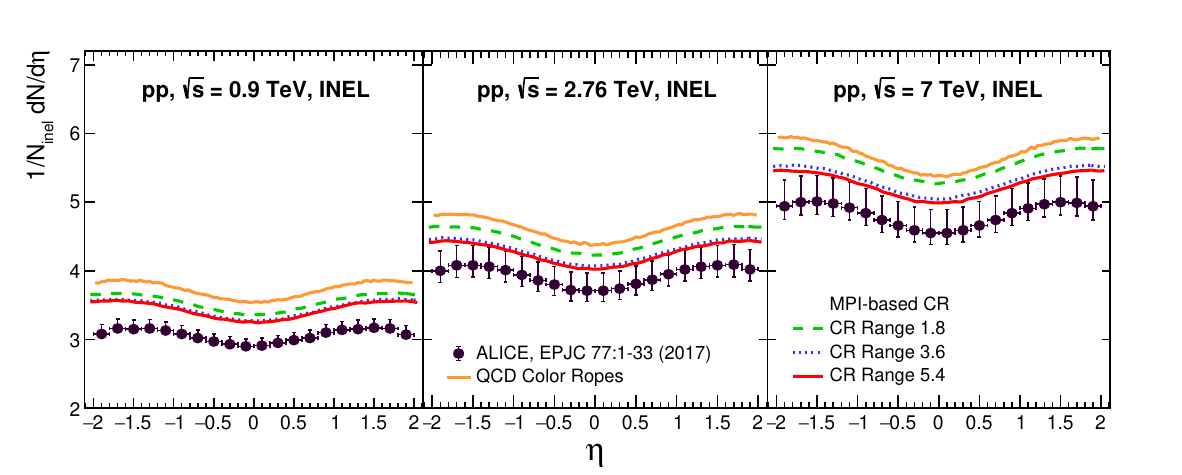}
    \includegraphics[width=\linewidth]{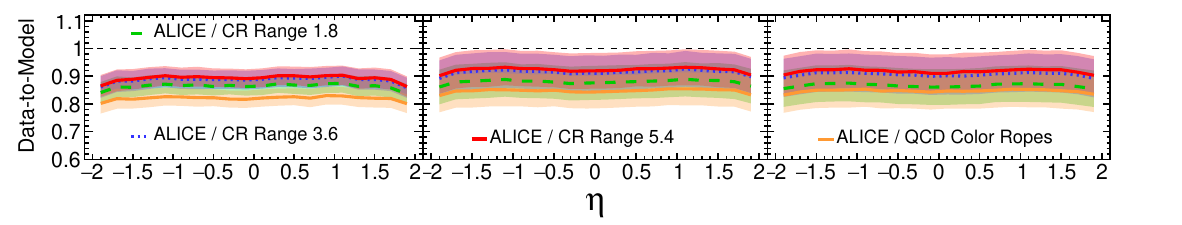}
    \caption{(Color online) Charged-particle multiplicities as a function of $\eta$ for INEL pp collisions at $\sqrt{s}$ = 0.9, 2.76 and 7 TeV simulated using \texttt{PYTHIA8} and plotted alongside ALICE measurements~\cite{adam2017}. ALICE data (\textit{magenta circles with systematic uncertainties}), CR Range 1.8 (\textit{green dashed line}), CR Range 3.6 (\textit{blue dotted line}), CR Range 5.4 (\textit{red solid line}), and QCD Color Ropes (\textit{orange solid line}). The \textit{bottom panels} show ratios of data and model calculations with uncertainties plotted in transparent bands.}
    \label{fig:eta_distributions}
\end{figure*}

\section{Methodology} \label{section4}

The data presented in this study were generated using the \texttt{PYTHIA8} Monte Carlo event generator (version 8.311), simulating pp collisions at center-of-mass energies of $\sqrt{s} = 0.9$, 2.76 and 7~TeV. For each of the collision energies, \textbf{50} Million inelastic (INEL) non-diffractive pp collisions were generated. Central to this study are four distinct observables: (a) the \textit{extensive} quantities - (i) FB multiplicity and (ii) FB summed transverse momentum ($\sum p_{\rm T}$) correlation strength; (b) the \textit{intensive} quantities - (iii) FB mean transverse momentum ($\overline p_{\rm T}$) correlation and (iv) the strongly intensive quantity ($\Sigma$). These observables were estimated using kinematics aligned with ALICE and ATLAS detector configurations, presented in Tab.~\ref{tab:kinematics}. The analysis is restricted to charged particles only, specifically $\pi^{\pm}$, $K^{\pm}$, $p$, and  $\overline p$. The central pseudorapidity region with $|\eta| < 0.8 $ (acceptance of the central barrel detector of ALICE) is sliced into symmetric forward ($\eta > 0$) and backward ($\eta< 0$) intervals, illustrated in Fig.~\ref{fig:illustrationnew}. The width of the $\eta$-windows were chosen as $\delta \eta =$ 0.2, 0.4, 0.6 and 0.8 for the calculation of $b_{\rm corr}(mult.)$, $b_{\rm corr}^{\overline p_{\rm T}}$ and $\Sigma_{\rm N_{\rm F} \rm N_{\rm B}}$. The value of $\delta \eta $ is chosen as 0.5 while estimating $b_{\rm corr}^{\sum p_{\rm T}}$. The FB multiplicity correlation analysis is further extended to encompass azimuthal and pseudorapidity-segmented correlations. The azimuthal plane is subdivided into 5 sectors, each of width $\delta\varphi = \pi/4$, illustrated in Fig.~\ref{fig:bcorr_nsep} (\textit{bottom}). Emphasizing correlations linked to soft particle production, the \texttt{PYTHIA8} events were simulated with the flag ``\texttt{SoftQCD:all = on}''. Further, enabling/disabling ISR and FSR through the ``\texttt{PartonLevel:ISR}" \& ``\texttt{PartonLevel:FSR}" switches have been done in order to isolate their contributions to FB correlation strength. Furthermore, other variants of \texttt{PYTHIA8} events were simulated by switching on ``QCD Color Ropes". The parameters governing QCD Color Ropes within the QCD-based CR scheme of \texttt{PYTHIA8} are provided in Tab.~\ref{tab:parameters}.


\begin{figure*}
    \centering
    \includegraphics[width=\linewidth]{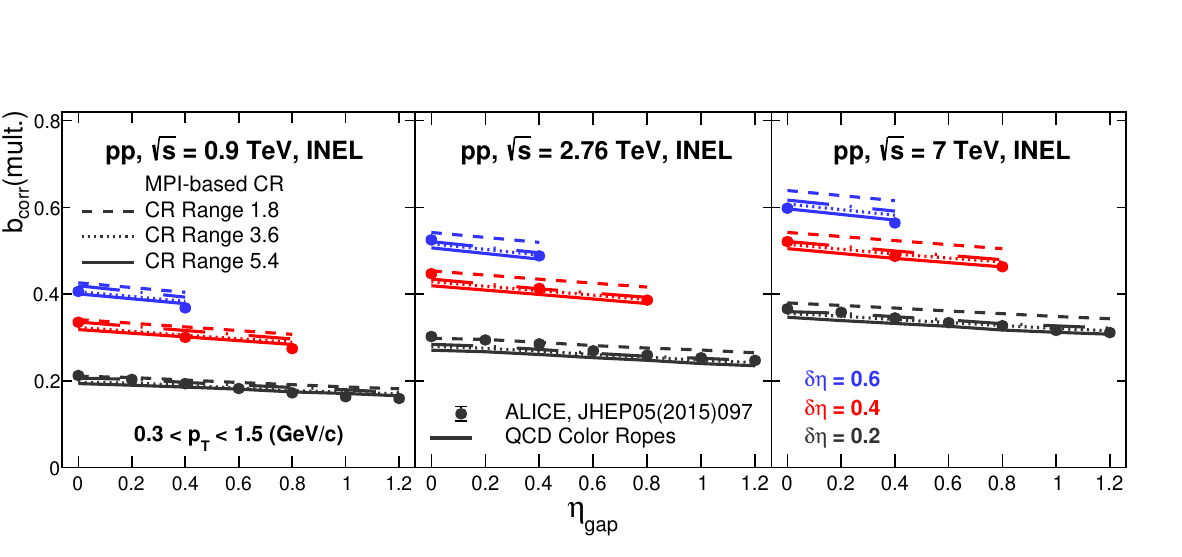}
     \includegraphics[width=1.\linewidth]{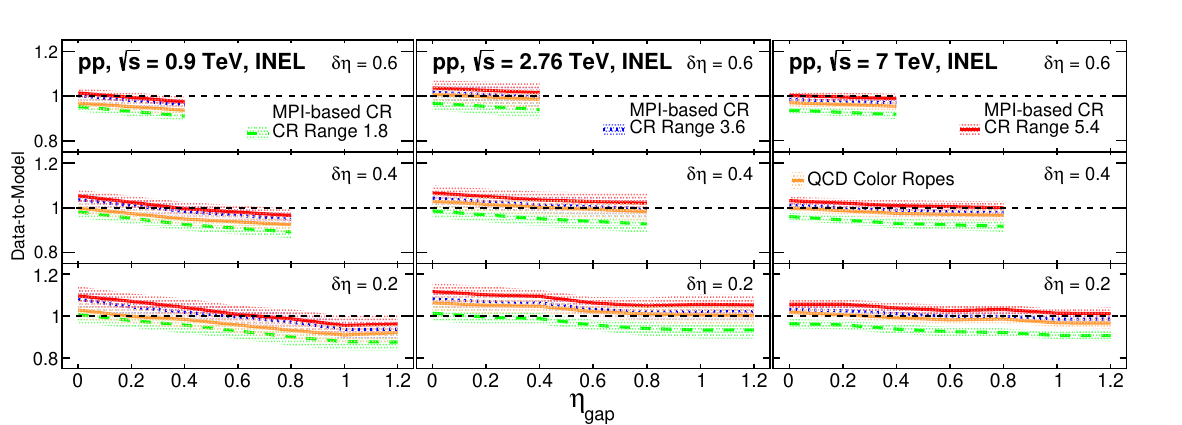}
    \caption{(Color online) FB multiplicity correlation strength, $b_{\rm corr}(mult.)$ in symmetrically opposite intervals ($\delta\eta$) plotted as a function of $\eta_{\rm gap}$ (\textit{top}) for INEL pp collisions at $\sqrt{s}$ = 0.9, 2.76 and 7 TeV, wherein \texttt{PYTHIA8} model calculations have been compared with ALICE results~\cite{adam2015}. The \textit{bottom panels} represent data-to-model ratios with uncertainties shown in bands.}
    \label{fig:bcorr_etagap}
\end{figure*}
\begin{figure*}
    \centering
    \includegraphics[width=\linewidth]{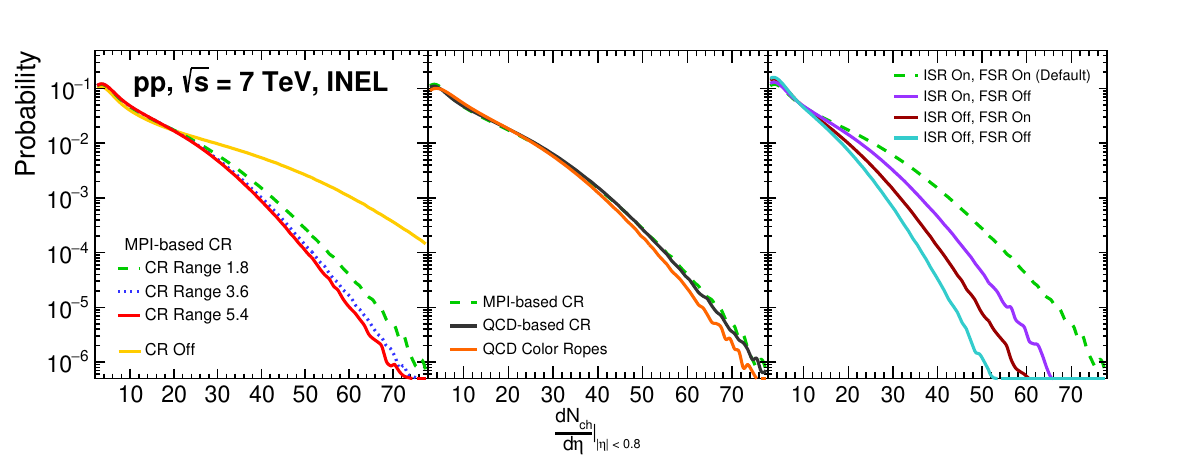}
    \caption{(Color online) Mid-rapidity charged-particle multiplicity yield in various settings and variants of \texttt{PYTHIA8} framework for INEL pp collisions at $\sqrt{s}$ = 7 TeV.}
    \label{fig:nch_distribution}
\end{figure*}

\section{Results and discussion} \label{section5}

In this work, we investigate forward-backward correlations of charged-particle multiplicity, summed transverse momentum, mean transverse momentum, and the strongly intensive quantity in pp collisions at $\sqrt{s} =$ 0.9, 2.76, and 7~TeV within the \texttt{PYTHIA8} (version 8.311) framework. The analysis has been carried out on non-diffractive inelastic (INEL) events using different variants of the \texttt{PYTHIA8} model, and the results were then compared to published  ATLAS~\cite{aad2012atlas} and ALICE~\cite{adam2015} results. To validate the generated data, pseudorapidity distributions of charged particles from each \texttt{PYTHIA8} tune were compared with experimental results. Fig.~\ref{fig:eta_distributions} (\textit{top}) presents the pseudorapidity distribution of charged particles at $\sqrt{s} =$ 0.9, 2.76 and 7 TeV as obtained with \texttt{PYTHIA8}. The curves compare three MPI-based color reconnection (CR) ranges (1.8, 3.6, and 5.4) and the QCD Color Rope scenario with ALICE data~\cite{adam2017}. The results show that all variants of the \texttt{PYTHIA8} model systematically overpredict experimental measurements, with the QCD Color Ropes mechanism showing the largest discrepancy. Notably, the CR range parameter of 5.4 achieves the best consistency with experimental results across all studied energies, a conclusion supported by the ratio comparison in Fig.~\ref{fig:eta_distributions} (\textit{bottom}). Given that pseudorapidity spectra exhibit sensitivity to the CR range and hadronization model, it is reasonable to expect that forward-backward (FB) correlations will similarly depend on these \texttt{PYTHIA8} parameters. Hence, an effort has been made to study the effect of CR and hadronization mechanisms on FB correlations.


\subsection{Role of CR Ranges and QCD Color Ropes on FB Multiplicity Correlation}

\subsubsection{$\eta_{\rm gap}$ dependence of $b_{\rm corr}(mult.)$}

FB multiplicity correlation strength, $b_{\rm corr}(mult.)$ as described by Eqn.~\ref{eq2} is plotted as a function of pseudorapidity gap ($\eta_{\rm gap}$) for four different pseudorapidity window widths ($\delta\eta$ = 0.2, 0.4, and 0.6) in Fig.~\ref{fig:bcorr_etagap} (\textit{top}). Both figures correspond to center-of-mass energies of $\sqrt{s}$ = 0.9, 2.76, and 7 TeV, with results obtained using various tunes of the \texttt{PYTHIA8} model. For the first time, the QCD Color Ropes mechanism present in \texttt{PYTHIA8} has been implemented to study FB correlations. The model calculations in both cases are compared with experimental data from ALICE collaboration~\cite{adam2015}. The figure clearly shows that, for both ALICE data and the \texttt{PYTHIA8} model, $b_{\rm corr}(mult.)$ exhibits a decreasing trend with increasing  $\eta_{\rm gap}$ irrespective of $\delta\eta$ and the CR range. This observation can be understood from the fact that the Short Range Correlations (SRCs) contributing to the correlation strength are only effective over a limited pseudorapidity range. Another observation in Fig.~\ref{fig:bcorr_etagap} (\textit{top}) is the increase of $b_{\rm corr}(mult.)$ with pseudorapidity window width ($\delta\eta$). These results are consistent with the extensive character of the $b_{\rm corr}(mult.)$ observable. Furthermore, a pronounced sensitivity of $b_{\rm corr}(mult.)$ to the CR range is observed, with $b_{\rm corr}(mult.)$ exhibiting a monotonic decrease as the CR range increases. This behavior reflects the dependence of forward-backward (FB) correlation strength on the particle multiplicity at mid-rapidity. As illustrated in Fig.~\ref{fig:nch_distribution} (\textit{left panel}), the mid-rapidity charged-particle yield $(dN_{\rm ch}/d\eta_{|\eta| < 0.8})$ decreases with increasing CR range. A better agreement of the \texttt{PYTHIA8} model results with the experimental mid-rapidity multiplicity leads to an improved description of the FB correlation strength when compared with the data. \\

The \textit{bottom panel} of Fig.~\ref{fig:bcorr_etagap} shows ratios between experimentally measured $b_{\rm corr}(mult.)$ values~\cite{adam2015} and the \texttt{PYTHIA8} calculations, within uncertainties. It is evident from the figures that the default \texttt{PYTHIA8} with CR Range 1.8 fails to explain experimental results in all the studied energies, with only a few isolated exceptions. However, the other CR Ranges (3.6 and 5.4) have been able to quantitatively explain the experimental data within uncertainty. The above observations can be understood by noting that the charged-particle yields corresponding to CR ranges of 3.6 and 5.4 quantitatively reproduce the experimental data, while the default CR range of 1.8 fails to do so. In contrast, the QCD Color Rope mechanism, despite significantly overestimating the overall charged-particle production (as evident from  Fig.~\ref{fig:eta_distributions}), appears to describe the data at  $\sqrt{s}$ = 2.76 and 7 TeV within experimental uncertainties. This seemingly good agreement is unexpected and cannot be attributed to the extensive nature of the $b_{\rm corr}(mult.)$ observable. Therefore, it underscores the need to explore intensive and strongly intensive observables for a more robust understanding of particle production dynamics.

\begin{figure*}
    \centering
    \includegraphics[width=1.\linewidth]{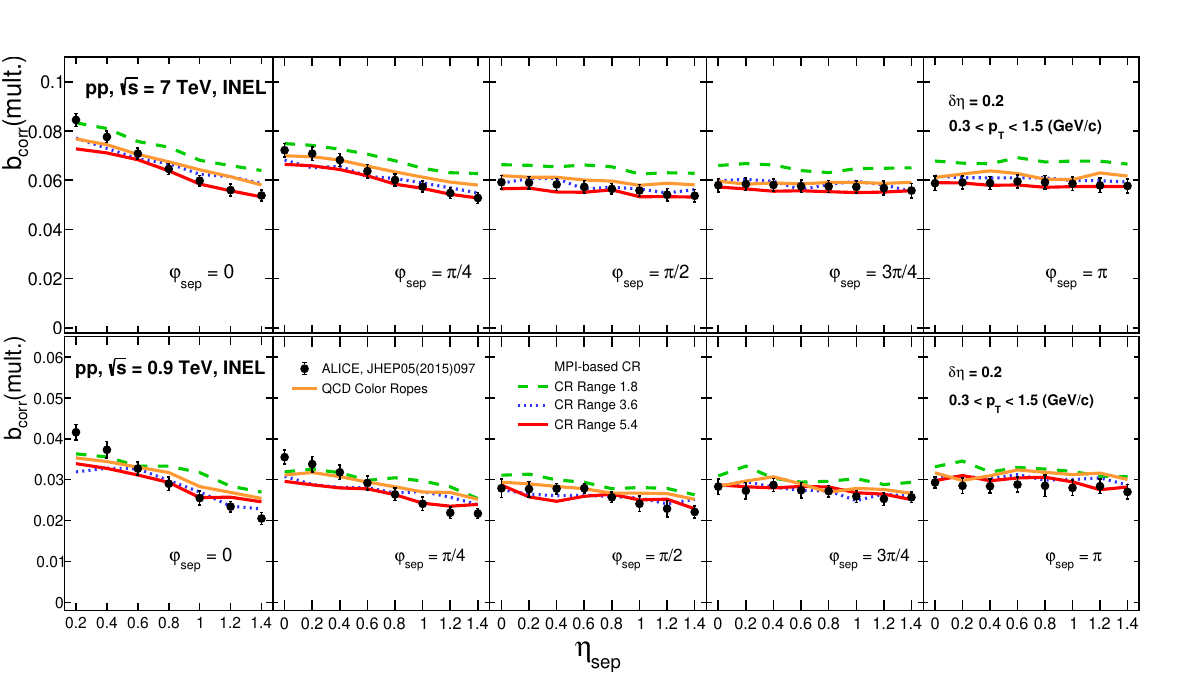}
    \includegraphics[width=1.\linewidth]{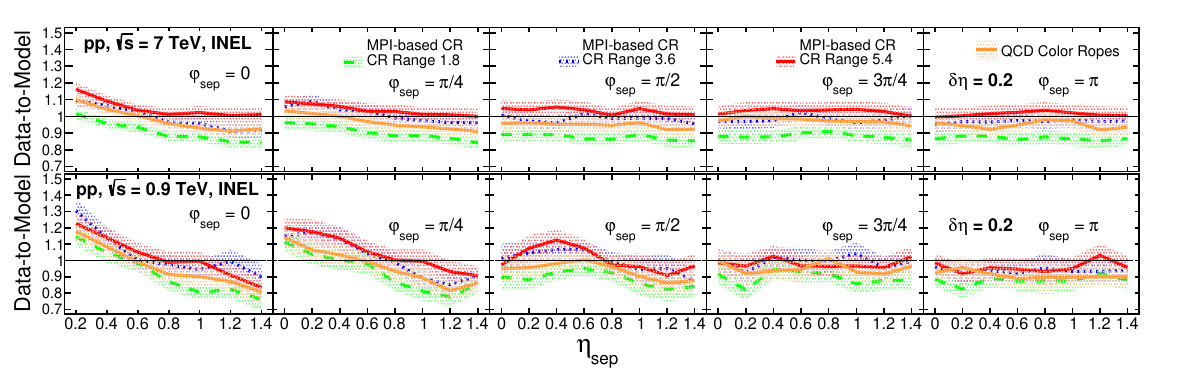}
    \includegraphics[width=0.65\linewidth]{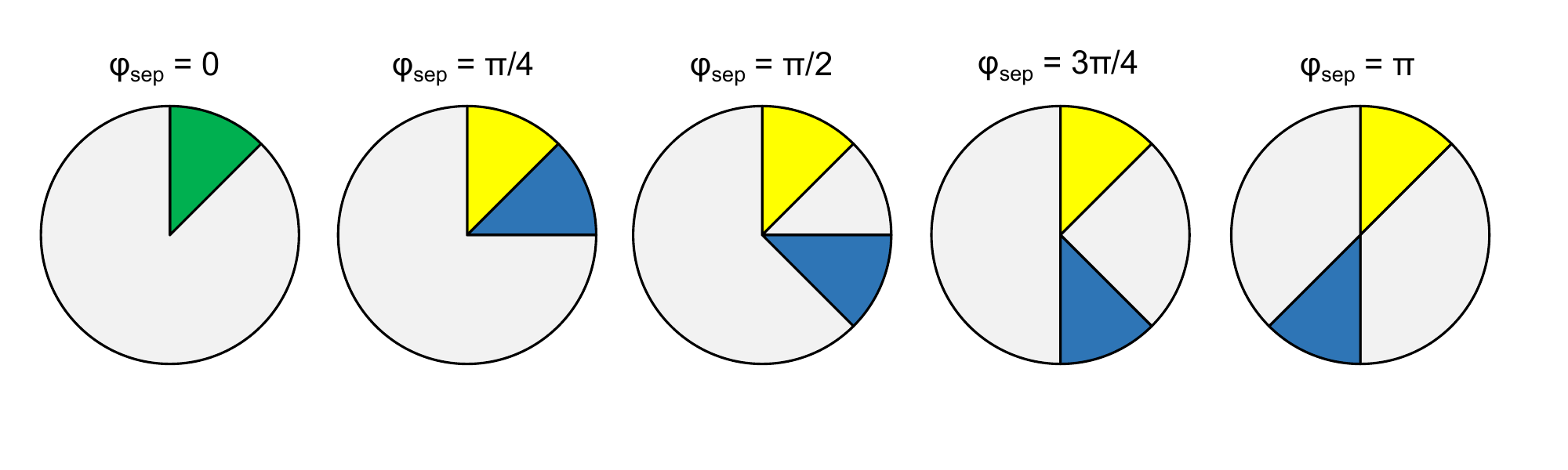}
    \caption{(Color online) FB multiplicity correlation strength, $b_{\rm corr}(mult.)$ for azimuthally separated $\eta$-$\varphi$ windows of $\varphi_{\rm sep}$ = 0, $\pi$/4, $\pi$/2, 3$\pi$/4 and $\pi$, plotted as a function of $\eta_{\rm sep}$ (\textit{top panels}) for \texttt{PYTHIA8} simulated INEL pp collisions at $\sqrt{s}$ = 0.9 and 7 TeV and compared with ALICE results~\cite{adam2015}. Data-to-model ratios are illustrated in the \textit{mid-panel} with uncertainties shown as bands. The \textit{bottom figure} houses an illustrative diagram showing five different azimuthal sectors.}
    \label{fig:bcorr_nsep}
\end{figure*}
\begin{figure*}
    \centering
    \includegraphics[width=0.9\linewidth]{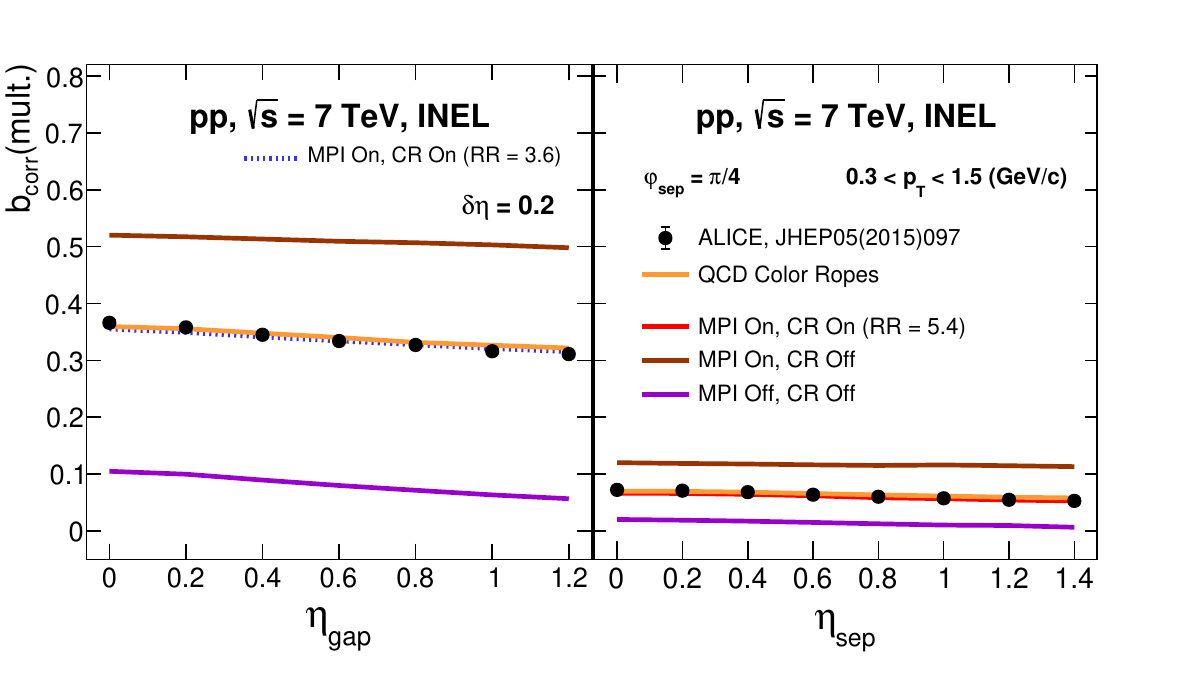}
    \caption{(Color online) FB multiplicity correlation strength, $b_{\rm corr}(mult.)$ plotted as a function of $\eta_{\rm gap}$ (\textit{left}) and azimuthally separated $\eta$-$\varphi$ window of $\varphi_{\rm sep}$ = $\pi$/4 (\textit{right}) with different combinations of MPI and CR for \texttt{PYTHIA8} generated INEL pp collisions at $\sqrt{s}$ = 7 TeV. ALICE results~\cite{adam2015} have been plotted alongside model calculations.}
    \label{fig:bcorr_ngap_nsep_cr_mpi_qcdrope}
\end{figure*}
\begin{figure}
    \centering
    \includegraphics[scale = 0.45]{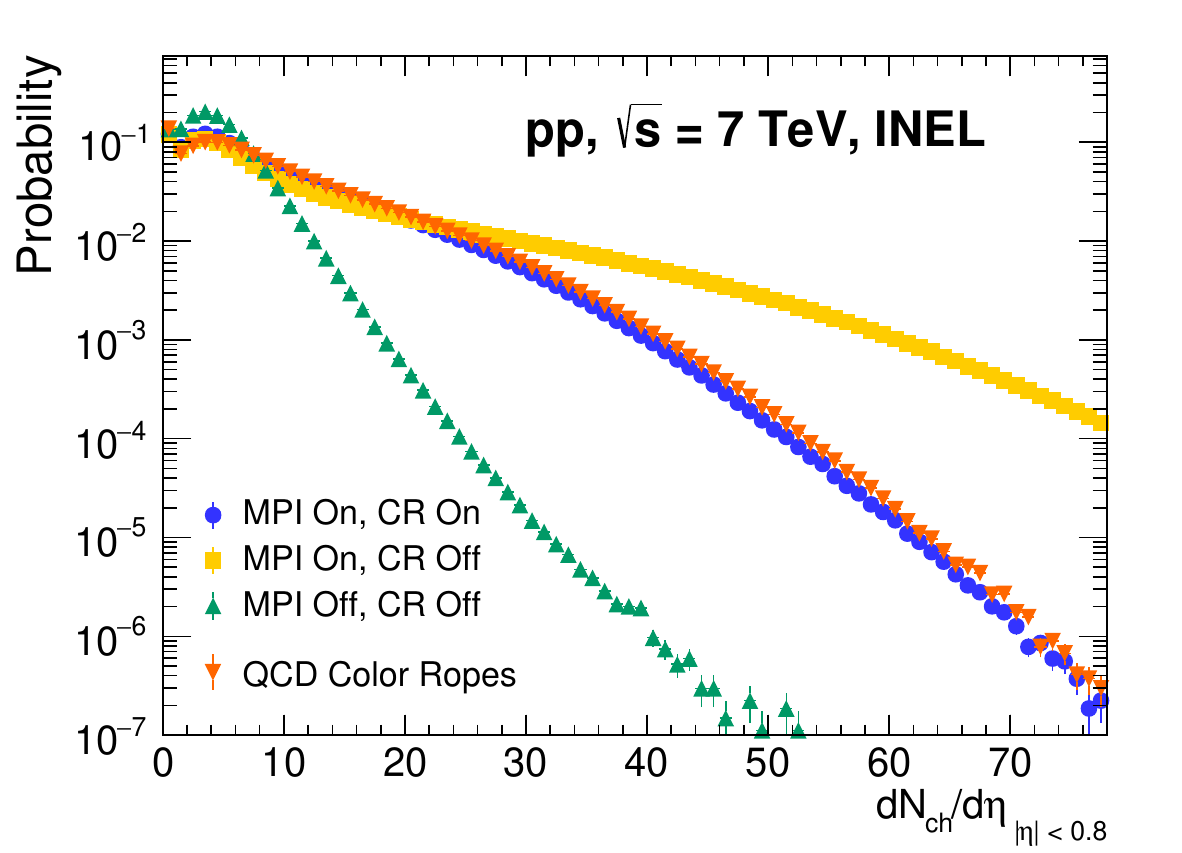}
    \caption{(Color online) Mid-rapidity charged-particle multiplicity ($\rm N_{\rm ch}$) yields for \texttt{PYTHIA8} simulated INEL pp collisions at $\sqrt{s}$ = 7 TeV in different MPI and CR settings.}
    \label{fig:nch_distribution_MPICR}
\end{figure}

\subsubsection{$\eta-\varphi$ dependence of $b_{\rm corr}(mult.)$}

Fig.~\ref{fig:bcorr_nsep} (\textit{top}) illustrates the dependence of the FB correlation strength, $b_{\rm corr}(mult.)$ on pseudorapidity separation, $\eta_{\rm sep}$, in pp collisions simulated with different variants of \texttt{PYTHIA8}~\cite{bierlich2022} at center-of-mass energies of $\sqrt{s}$ = 0.9 and 7 TeV and compared to ALICE data~\cite{adam2015}. The analysis systematically explores five distinct azimuthal separations, $\varphi_{\rm sep}$: 0, $\pi/4$, $\pi/2$, $3\pi/4$ and $\pi$, as shown in Fig.~\ref{fig:bcorr_nsep} (\textit{bottom}), where the transverse plane is divided into five azimuthal sectors, to disentangle the contributions from SRCs and LRCs among produced particles. It is seen from the present study that at both collision energies, a pronounced decline in $b_{\rm corr}(mult.)$ with increasing $\eta_{\rm sep}$ is observed for azimuthally adjacent sectors ($\varphi_{\rm sep}$ = 0 and $\pi/4$). This observation can be attributed to the dominance of SRCs among particles concentrated in the narrow phase space separated by $\varphi_{\rm sep} \leq \pi/4$, which are believed to arise from localized interactions (e.g., jet fragmentation, decays of clusters, etc.). At $\varphi_{\rm sep}$ = $\pi/2$ and beyond, $b_{\rm corr}(mult.)$ plateaus, reaching saturation values of approximately 0.06 and 0.03 for $\sqrt{s}$ = 7 and 0.9 TeV, respectively. This saturation signifies the transition to LRC-dominated regimes. Tuned \texttt{PYTHIA8} (with CR Range = 1.5, $p_{\rm T_{\rm 0}}^{\rm ref} = 2.6$ GeV/c, and $\langle N_{\rm MPI} \rangle = 2.7$) framework was previously deployed~\cite{kundu2019} to explain the $\eta-\varphi$ dependence of $b_{\rm corr}(mult.)$. But it could not adequately replicate the experimental observations. In the present study, an attempt has therefore been made to use different tunes of \texttt{PYTHIA8} model. It is evident from Fig.~\ref{fig:bcorr_nsep} (\textit{middle}) that the CR range 1.8 fails to quantitatively reproduce ALICE data across all azimuthal sectors, underscoring its inadequacy in modeling both SRCs and LRCs. In contrast, the CR range 5.4 provides a satisfactory description of the $\sqrt{s}$ = 7 TeV results, while the CR range 3.6, 5.4, and QCD Color Ropes mechanism align with $\sqrt{s}$ = 0.9 TeV results, as validated by the ratio plot shown in Fig.~\ref{fig:bcorr_nsep} (\textit{middle}). It is to be noted that the \texttt{QGSM} model, based on the Regge-theory approach where fluctuations in the number of cut pomerons are explicitly considered, could explain the ALICE results at higher pseudorapidity gap~\cite{bravina2018}.


\subsection{Effect of MPI, CR \& QCD Color Ropes on FB Multiplicity Correlation}

Fig.~\ref{fig:bcorr_ngap_nsep_cr_mpi_qcdrope} (\textit{left}) represents $b_{\rm corr}(mult.)$ as a function of $\eta_{\rm gap}$ for four different scenarios in \texttt{PYTHIA8}: (i) MPI On, CR On (ii) MPI On, CR Off (iii) MPI Off, CR Off and (iv) QCD Color Ropes. The ALICE data~\cite{adam2015} align closely with \texttt{PYTHIA8} predictions only when both MPI and CR are enabled, underscoring the importance of these initial and final state phenomena. It is further seen from Fig.~\ref{fig:bcorr_ngap_nsep_cr_mpi_qcdrope} that the QCD Color Rope mechanism provides a better quantitative description of the ALICE results. With MPI disabled mode (``MPI Off"), the correlation strength drops drastically compared to the MPI-enabled case. However, disabling CR in MPI-enabled scenarios (``MPI On, CR Off") overestimates $b_{\rm corr}(mult.)$ results from ALICE. A comparable trend is observed while analyzing $b_{\rm corr}(mult.)$ as a function of $\eta_{\rm sep}$ under the additional constraint of $\varphi_{\rm sep} = \pi/4$, as illustrated in Fig.~\ref{fig:bcorr_ngap_nsep_cr_mpi_qcdrope} (\textit{right}). The trends seen in Fig.~\ref{fig:bcorr_ngap_nsep_cr_mpi_qcdrope} can be understood through Fig.~\ref{fig:nch_distribution_MPICR}, which presents the mid-rapidity charged-particle distributions ($dN_{\rm ch}/d\eta_{|\eta| < 0.8}$) plotted for different \texttt{PYTHIA8} configurations. Fig.~\ref{fig:nch_distribution_MPICR} reveals that $dN_{\rm ch}/d\eta_{|\eta| < 0.8}$ in the MPI On, CR Off and MPI Off, CR Off configurations exhibit higher and lower yields, respectively, compared to the QCD Color Ropes and the scenario where both MPI and CR are enabled. The above results are consistent with the extensive nature of $b_{\rm corr}(mult.)$, which exhibits the trivial volume dependence.

\begin{figure}
    \centering
    \includegraphics[width=1.\linewidth]{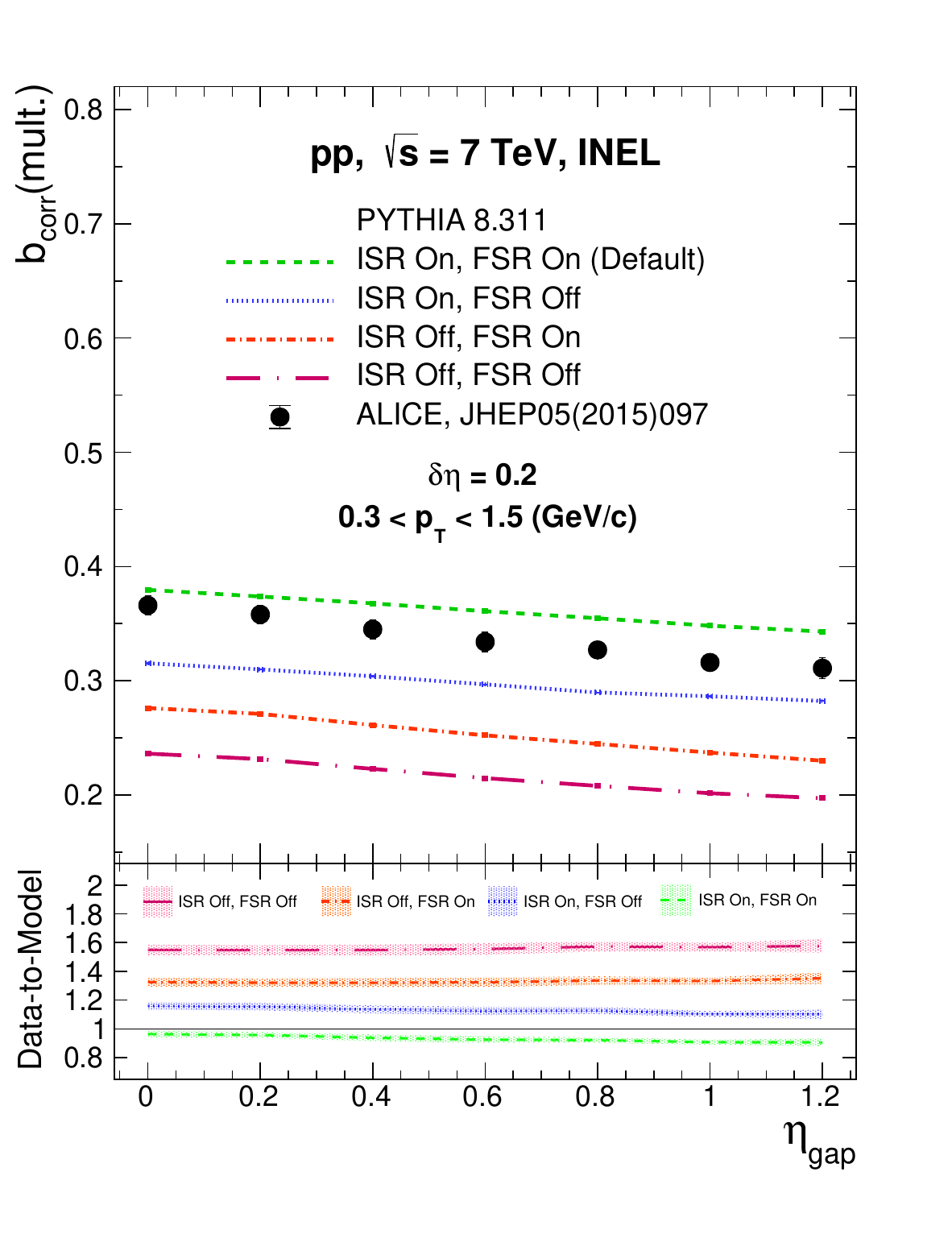}
    \caption{(Color online) FB multiplicity correlation strength, $b_{\rm corr}(mult.)$ for $\sqrt{s}$ = 7 TeV, plotted as a function of $\eta_{\rm gap}$ for \texttt{PYTHIA8} simulated INEL pp collisions. The \textit{bottom pad} shows a ratio plot between ALICE~\cite{adam2015} and model results, and uncertainties are shown in the shaded portion.}
    \label{fig:bcorr_etagap_isrfsr}
\end{figure}

\subsection{Impact of Parton Showers on FB Multiplicity Correlation}

\begin{figure}
    \centering
    \includegraphics[scale=0.45]{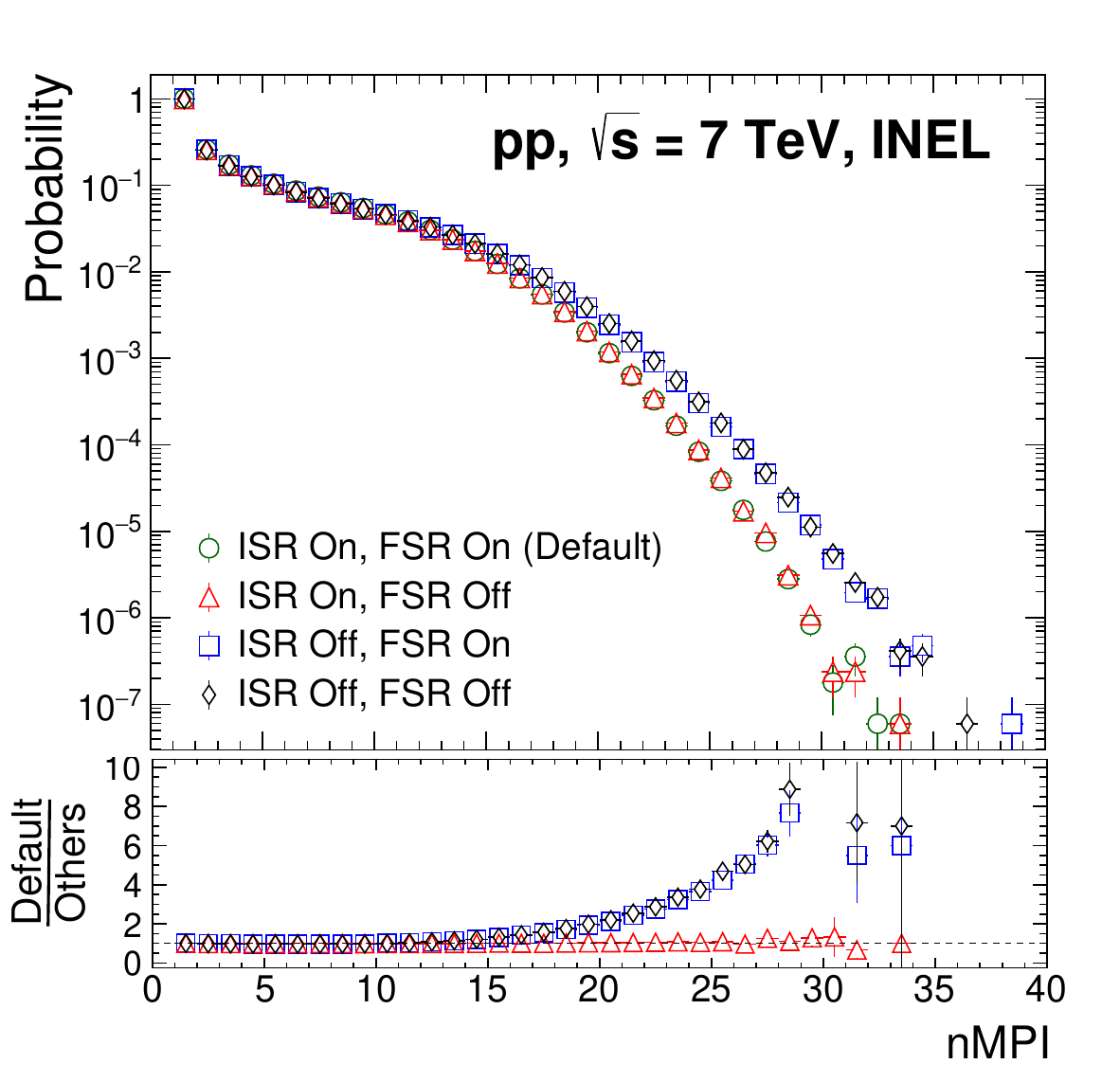}
    \caption{(Color online) Probability distribution of nMPI, plotted for different combinations of ISR \& FSR for \texttt{PYTHIA8} generated INEL pp collisions at $\sqrt{s}$ = 7 TeV. The \textit{bottom pad} shows ratio between the default case and other configurations.}
    \label{fig:mpinch_distribution_isrfsr}
\end{figure}

Initial and Final State Radiations (ISR \& FSR) are essential `hard' pQCD processes that can significantly impact various observables in high-energy particle collisions~\cite{ortiz2021,bencedi2021}. Through this study, we aim to elucidate the relative importance of ISR and FSR in shaping correlations in pp collisions at the soft QCD regime ($0.3 < p_{\rm T} < 1.5$ GeV/c). By definition, ISR refers to the emission of gluons or photons by the incoming partons before the hard scattering process. Similarly, the radiations after the hard scattering are often termed as FSR. In Fig.~\ref{fig:bcorr_etagap_isrfsr} (\textit{upper}), $b_{\rm corr}(mult.)$ is plotted as a function of $\eta_{\rm gap}$ under four configurations: (i) ISR On, FSR On (default) (ii) ISR On, FSR Off (iii) ISR Off, FSR On and (iv) ISR Off, FSR Off and compared to ALICE results~\cite{adam2015}. The default scenario serves as the baseline for comparison, while the alternate configurations isolate the effects of ISR and FSR. The \textit{lower panel} of Fig.~\ref{fig:bcorr_etagap_isrfsr} quantifies deviations between experimental results and model predictions via a ratio plot with uncertainty bands. Key findings reveal that the ISR-disabled configurations have the maximum impact on correlation strength. The significant reduction in the correlation strength for the ISR-disabled case can be understood from the probability distribution of charged-particle multiplicity at mid-rapidity ($dN_{\rm ch}/d\eta_{|\eta| < 0.8}$), which shows the lowest average charged-particle multiplicity when plotted alongside other configurations (\textit{lower panel},~Fig.~\ref{fig:mpinch_distribution_isrfsr}). In contrast, FSR, a final state phenomenon, has a relatively smaller impact on the estimated correlation strength. \\

In order to further validate the relative contribution of ISR and FSR on $b_{\rm corr}(mult.)$, we have plotted the probability distribution of the number of multi-parton interactions (nMPI) in Fig.~\ref{fig:mpinch_distribution_isrfsr} for the four scenarios mentioned above. MPIs are initial-state processes driven by parton density in the incoming protons. It is seen from Fig.~\ref{fig:mpinch_distribution_isrfsr} that the enabling or disabling of FSR virtually has a negligible effect on nMPI, which is also validated from the ratio plot. 
On the other extreme, an inverse correlation is observed between the number of multiple parton interactions (nMPI) and the presence of initial state radiation (ISR). Specifically, the mean nMPI, $\langle \rm nMPI \rangle$, increases when ISR is turned off.
This behavior can be attributed to the following considerations: (a) ISR removes energy from the incoming partons prior to the hard scattering, effectively increasing the Bjorken‑$x$ values; (b) this increase in Bjorken‑$x$ leads to a reduction in the parton density; and (c) the decreased parton density, in turn, suppresses the likelihood of multiple parton interactions, thereby lowering $\langle \rm nMPI \rangle$. \\

The decrease in the magnitude of $b_{\rm corr}(mult.)$ in ISR-off configurations—despite higher nMPI—stems from energy redistribution: retained initial-state energy increases partonic interaction probability (via Bjorken-$x$ dynamics), but dilutes energy per MPI. This reduces $dN_{\rm ch}/d\eta_{|\eta| < 0.8}$ and suppresses event-by-event multiplicity fluctuations (Fig.~\ref{fig:mpinch_distribution_isrfsr}). In contrast, FSR only moderates $dN_{\rm ch}/d\eta_{|\eta| < 0.8}$ via final-state fragmentation, leaving nMPI—and thus initial-state dynamics—unperturbed. Thus, $b_{\rm corr}(mult.)$ is primarily sensitive to ISR-mediated MPI energy loss, not MPI count alone.

\begin{figure*}
    \centering
    \includegraphics[width=\linewidth]{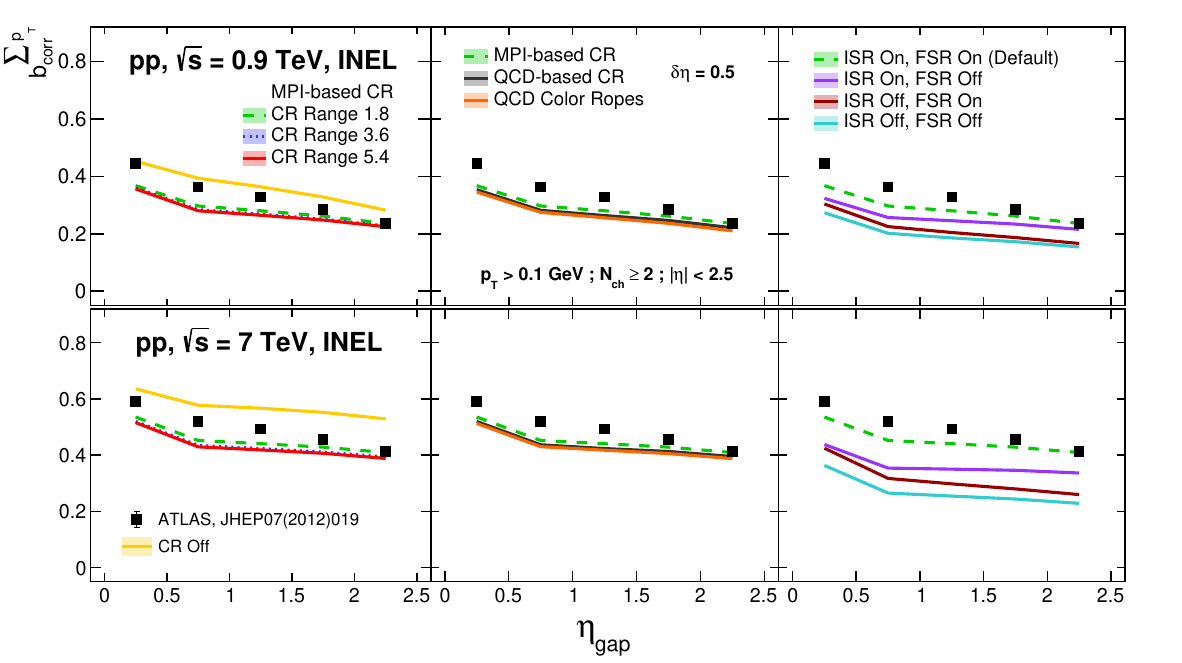}
    \caption{(Color online) FB summed transverse momentum correlation, $b_{\rm corr}^{\sum p_{\rm T}}$ plotted as a function of $\eta_{\rm gap}$ for \texttt{PYTHIA8} generated INEL pp collisions at $\sqrt{s}$ = 0.9 and 7 TeV alongside ATLAS results~\cite{aad2012atlas}. Dependence of CR Ranges (\textit{left column}), CR Mechanisms (\textit{middle column}), and Parton Shower (\textit{right column}) have been quantified in subsequent plots. The quadrature sum of systematic and statistical uncertainties has been plotted in the ATLAS data.}
    \label{fig:bcorr_summedpT}
\end{figure*}
\begin{figure*}
   \centering
    \includegraphics[width=\linewidth]{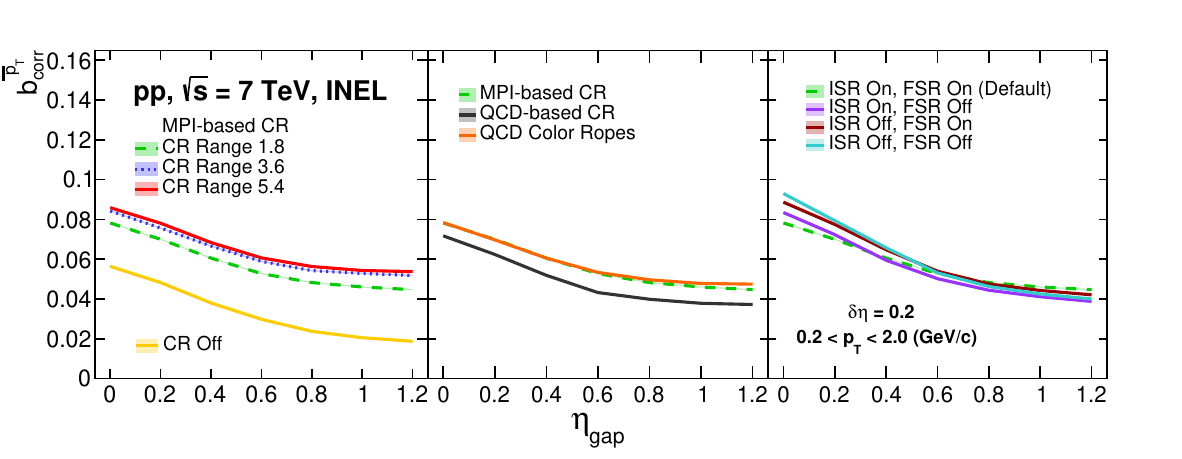}
    \caption{(Color online) FB mean transverse momentum correlation, $b_{\rm corr}^{\overline p_{\rm T}}$ plotted as a function of $\eta_{\rm gap}$ for distinct CR ranges of MPI-based CR (\textit{left}), different hadronization mechanisms in terms of CR and Ropes (\textit{middle}), and four combinations of QCD radiations (\textit{right}) for INEL pp collisions at $\sqrt{s}$ = 7 TeV simulated using \texttt{PYTHIA8} model. The statistical uncertainties are within the bands as shown.}
    \label{fig:bcorr_etagap_meanpT}
\end{figure*}
\begin{figure*}
    \centering
    \includegraphics[width=\linewidth]{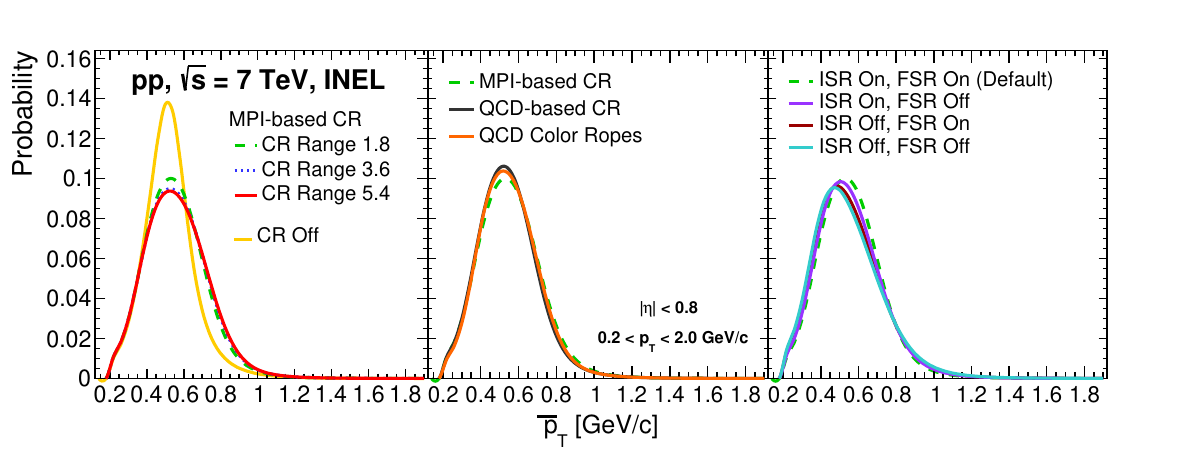}
    \caption{(Color online) Event-by-event $\overline p_{\rm T}$ distribution, plotted for pp collisions at $\sqrt{s}$ = 7 TeV for different settings and models of \texttt{PYTHIA8} framework.}
    \label{fig:probability_MeanpT}
\end{figure*}

\subsection{FB summed $p_{\rm T}$ correlation: Role of CR Ranges, CR Mechanisms and Parton Showers}

In this section, we explore another extensive quantity, the FB summed transverse momentum correlation denoted by $b_{\rm corr}^{\sum p_{\rm T}}$, plotted as a function of $\eta_{\rm gap}$ in Fig.~\ref{fig:bcorr_summedpT} for pp collisions simulated using \texttt{PYTHIA8} at its various settings alongside ATLAS data~\cite{aad2012atlas}. It can be seen from this figure (Fig.~\ref{fig:bcorr_summedpT}, \textit{left panels}) that the \texttt{PYTHIA8} variants qualitatively match the decreasing trend of the experimental data at all the studied energies. However, the magnitude of the quantitative discrepancy diminishes progressively with increasing $\eta_{\rm gap}$. Furthermore, while the correlation strength exhibits a weak dependence on the CR range, the discrepancy between the model predictions and experimental data tends to increase as the CR strength is enhanced. This observation is consistent with Fig.~\ref{fig:nch_distribution}, which shows that the mid-rapidity charged particle yield varies minimally with CR and decreases slightly as the CR range increases. This finding is consistent with the extensive nature of this observable, whose magnitude scales with the system volume where fluctuations originate. \\

In the \textit{middle panels} of Fig.~\ref{fig:bcorr_summedpT}, $b_{\rm corr}^{\sum p_{\rm T}}$ is plotted against $\eta_{\rm gap}$ for different CR models such as MPI-based, perturbative-QCD–inspired, and QCD Color Rope model. Similar to the previous results, the summed-$p_{T}$ correlation strength is also found to be weakly dependent on the choice of CR mechanism, and among the three mechanisms tested, the MPI-based CR is most consistent with the ATLAS data. \\

In Fig.~\ref{fig:bcorr_summedpT} (\textit{right panels}), we test the parton shower dependence on $b_{\rm corr}^{\sum p_{\rm T}}$. In contrast to the minimal sensitivity observed for the CR range and CR models, the correlation strength is highly sensitive to parton showers.
Being an extensive quantity, the behavior is though similar to what was observed in Fig.~\ref{fig:bcorr_etagap_isrfsr}, with ISR having a more significant effect than FSR, clearly depicted by the $dN_{\rm ch}/d\eta_{|\eta| < 0.8}$ distribution in the \textit{right plot} of Fig.~\ref{fig:nch_distribution}. 

\begin{table*}
    \centering
    \caption{Standard deviation ($\sigma$) of the event-by-event $\overline p_{\rm T}$ distributions in different variants and settings of \texttt{PYTHIA8} framework ($|\eta| < 0.8$, $0.2 < p_{\rm T} < 2.0$ GeV/c). \\}
    \label{tab:meanpT_stddev}
    \begin{tabular}{cccccc}
        \hline
        \hline
        \noalign{\vskip 0.2em}
        \textbf{Setting} & \textbf{Std. Dev.} & \textbf{Variant} & \textbf{Std. Dev.} & \textbf{Setting} & \textbf{Std. Dev.} \\
        \noalign{\vskip 0.2em}
        \hline
        \hline
        \noalign{\vskip 0.3em}
        CR Off & 0.1444  & & & ISR On, FSR On & 0.1664 \\
        
        CR Range 1.8 & 0.1664 & MPI-based CR & 0.1664 & ISR On, FSR Off & 0.1746 \\
        
        CR Range 3.6 & 0.1715 & QCD-based CR & 0.1607 & ISR Off, FSR On & 0.1797 \\
        
        CR Range 5.4 & 0.1728 & QCD Color Ropes & 0.1605 & ISR Off, FSR Off & 0.1870 \\
        
        \noalign{\vskip 0.3em}
        \hline
        \hline
    \end{tabular}
    \begin{tablenotes}
\small
\centering
\item[] CR $\rightarrow$ Color Reconnection; ISR \& FSR $\rightarrow$ Initial and Final State Radiation; uncertainty $\sim 10^{-5}$ 
    \end{tablenotes}
\end{table*}
\begin{figure*}
    \centering
    \includegraphics[width=\linewidth]{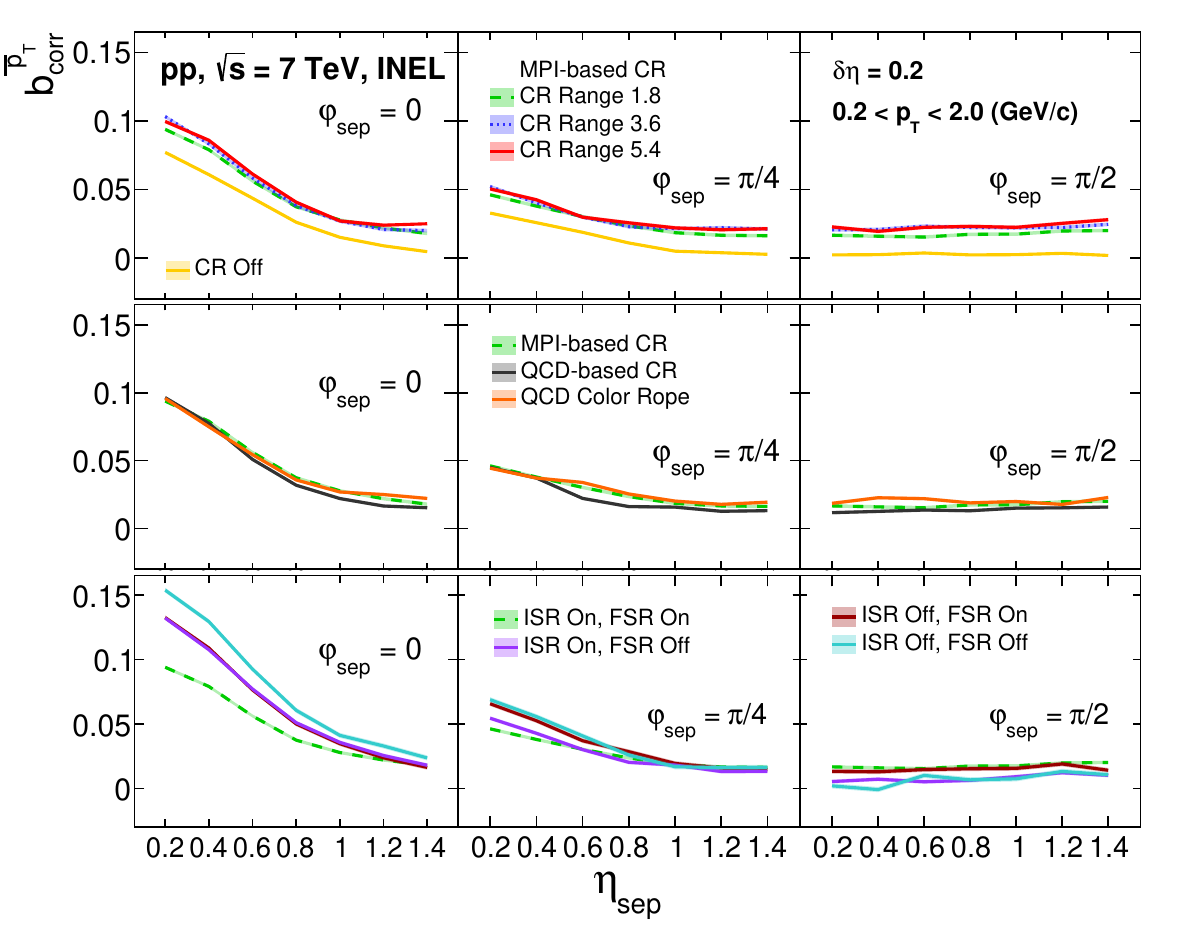}
    \caption{(Color online) FB $\overline p_{\rm T}$ correlation strength, $b_{\rm corr}^{\overline p_{\rm T}}$ for azimuthally separated $\eta$-$\varphi$ windows of $\varphi_{\rm sep}$ = 0, $\pi$/4 and $\pi$/2, plotted as a function of $\eta_{\rm sep}$ for \texttt{PYTHIA8} simulated INEL pp collisions at $\sqrt{s}$ = 7 TeV.}
    \label{fig:meanpT_azimuthal}
\end{figure*}
\begin{figure*}
    \centering
    \includegraphics[width=\linewidth]{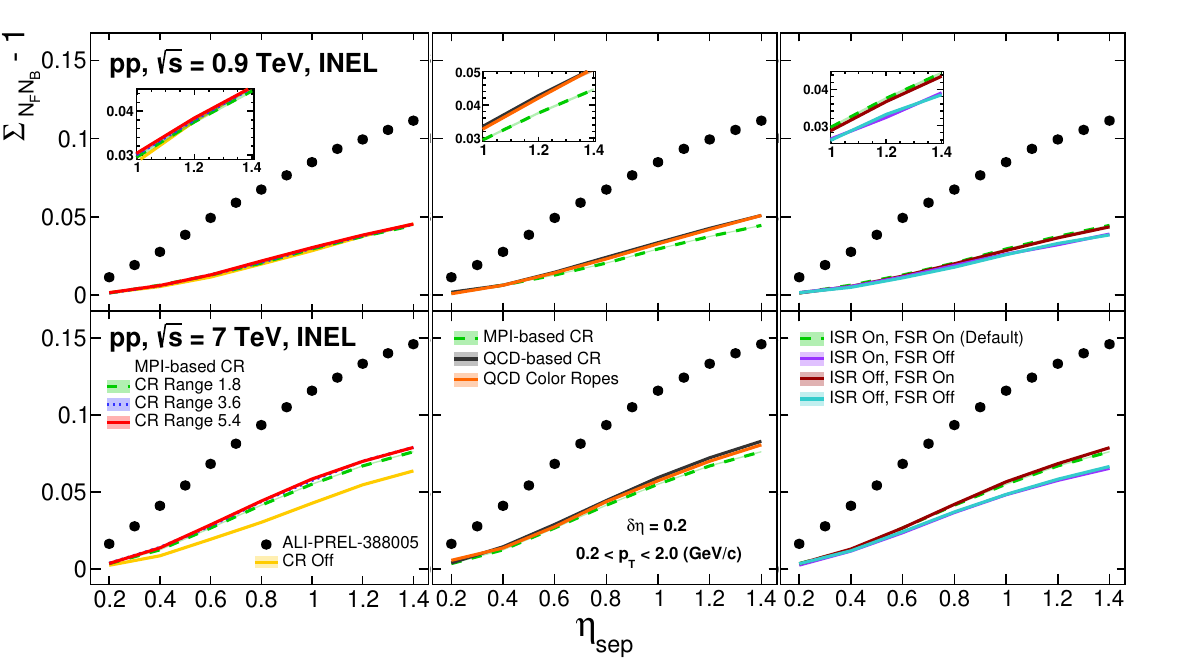}
    \caption{(Color online) Strongly intensive quantity minus one, $\Sigma_{\rm N_{\rm F} \rm N_{\rm B}} - 1$, plotted as a function of $\eta_{\rm sep}$ with $\delta\eta$ = 0.2. Dependence on CR Range (\textit{left column}), CR mechanisms (\textit{middle column}), and parton shower dependence (\textit{right column}) for INEL pp collisions simulated using \texttt{PYTHIA8} at $\sqrt{s}$ = 0.9 and 7 TeV. \textit{Solid markers} denote the ALICE Preliminary results~\cite{erokhin2021} and \textit{bands} represent the model calculations that include statistical uncertainties.}
    \label{fig:siq_all}
\end{figure*}

\subsection{FB mean $p_{\rm T}$ correlation: Roles of CR Ranges, CR Mechanisms and Parton Showers}

\subsubsection{$\eta_{\rm gap}$ dependence of $b_{\rm corr}^{\overline p_{\rm T}}$}

In Fig.~\ref{fig:bcorr_etagap_meanpT} (\textit{left}), the FB mean transverse momentum correlation, $b_{\rm corr}^{\overline p_{\rm T}}$, is plotted as a function of $\eta_{\rm gap}$ for different strengths of MPI-based CR along with the CR Off configuration at $\sqrt{s}$ = 7 TeV. A decreasing trend of $b_{\rm corr}^{\overline p_{\rm T}}$ with increasing $\eta_{\rm gap}$ is evident from the figure. Further, the correlation strength is found to decrease with decreasing CR range. The value of $b_{\rm corr}^{\overline p_{\rm T}}$ is seen to be the least at all $\eta_{\rm gap}$ for the CR off scenario. This observation is in contrast to what was seen in the case of both the FB multiplicity and summed transverse momentum correlations. A similar finding was also reported in Ref.~\cite{kundu2019}. This decreasing trend of the correlation strength can be attributed to the decrease of the standard deviation of event-by-event $\overline p_{\rm T}$ distribution with decreasing CR range as seen from Fig.~\ref{fig:probability_MeanpT} and Table~\ref{tab:meanpT_stddev}. \\

In Fig.~\ref{fig:bcorr_etagap_meanpT} (\textit{middle panel}), we test different hadronization schemes in terms of CR mechanisms such as MPI-based CR, QCD-based CR and QCD Color Ropes to assess their impact on the $b_{\rm corr}^{\overline p_{\rm T}}$. 
The results show that MPI‑based CR and Color Ropes produce similar correlation strengths, while the QCD-based CR model exhibits the weakest correlation strength among the three. This behavior arises because the QCD‑based reconnection scheme enforces strict SU(3) colour‑compatibility and only allows reconnections that minimize the string length~\cite{tumasyan2023cms}. This produces compact junctions with highly local reconnections and possibly suppresses long‑range correlation, yielding the smallest forward–backward $\overline p_{\rm T}$ correlations among the three options. By contrast, the default MPI‑based CR  reconnects partons across multiple MPI systems based primarily on kinematic proximity that minimizes the string length. This allows broader inter-MPI colour flow and results in higher  $\overline p_{\rm T}$ correlations than that of the QCD-inspired CR mechanism. Finally, enabling rope hadronization merges overlapping transverse strings into higher‑tension ``ropes", imparting collective transverse boosts might enhance the correlation strength to levels comparable to those found in MPI‑based CR. \\

In Fig.~\ref{fig:bcorr_etagap_meanpT} (\textit{right}), we have plotted $b_{\rm corr}^{\overline p_{\rm T}}$ vs $\eta_{\rm gap}$ for four different combinations of ISR and FSR using \texttt{PYTHIA8} to ascertain the effect of QCD radiations on the FB mean transverse momentum correlation. It is seen from the figure that the magnitude of the correlation is roughly independent of the QCD radiations, and this observation is in sharp contrast with the extensive observables such as FB multiplicity and summed $p_{\rm T}$ correlations.


\subsubsection{$\eta-\varphi$ dependence of $b_{\rm corr}^{\overline
p_{\rm T}}$}

We conducted a differential analysis to examine the dependence of the transverse momentum correlation, $b_{\rm corr}^{\overline p_{\rm T}}$, on azimuthal separation ($\varphi_{\rm sep}$), plotted against pseudorapidity separation ($\eta_{\rm sep}$) at a fixed $\delta\eta = 0.2$. This investigation was performed across three azimuthal sectors: $\varphi_{\rm sep} = 0$, $\pi/4$, and $\pi/2$. The study aimed to assess the influence of color reconnection (CR) range, CR mechanisms, and QCD radiation on $b_{\rm corr}^{\overline{p}_{\rm T}}$. For all \texttt{PYTHIA8} model configurations tested, we observed a consistent $\eta_{\rm sep}$ dependence of $b_{\rm corr}^{\overline{p}_{\rm T}}$ that mirrors the behavior previously seen in forward-backward multiplicity correlations. Specifically, at $\varphi_{\rm sep} = 0$, $b_{\rm corr}^{\overline{p}_{\rm T}}$ decreases sharply with increasing $\eta_{\rm sep}$, while larger azimuthal separations ($\varphi_{\rm sep} = \pi/4$ and $\pi/2$) show progressively weaker rates of decrease with $\eta_{\rm sep}$. \\

The \textit{top row} of Fig.~\ref{fig:meanpT_azimuthal}
shows how the CR range affects the transverse momentum correlation, $b_{\rm corr}^{\overline p_{\rm T}}$. Turning CR off consistently gives the lowest correlation values across all azimuthal directions, matching earlier findings in Fig.~\ref{fig:bcorr_etagap_meanpT}. While $b_{\rm corr}^{\overline p_{\rm T}}$ depends only weakly on CR range, it slightly increases when larger ranges are used. The \textit{middle row} compares different CR mechanisms, revealing that $b_{\rm corr}^{\overline p_{\rm T}}$
is nearly identical across CR models, though QCD-based CR shows marginally lower values across all the azimuthal sectors. Most importantly, the \textit{bottom row} demonstrates that $b_{\rm corr}^{\overline p_{\rm T}}$ is highly sensitive to QCD radiations (initial and final-state radiation) for $\varphi_{\rm sep}=0$. This sensitivity decreases with the increase of $\varphi_{\rm sep}$, indicating that QCD radiations play the dominant role in generating short-range correlations.

\begin{figure*}
    \centering
    \includegraphics[width=\linewidth]{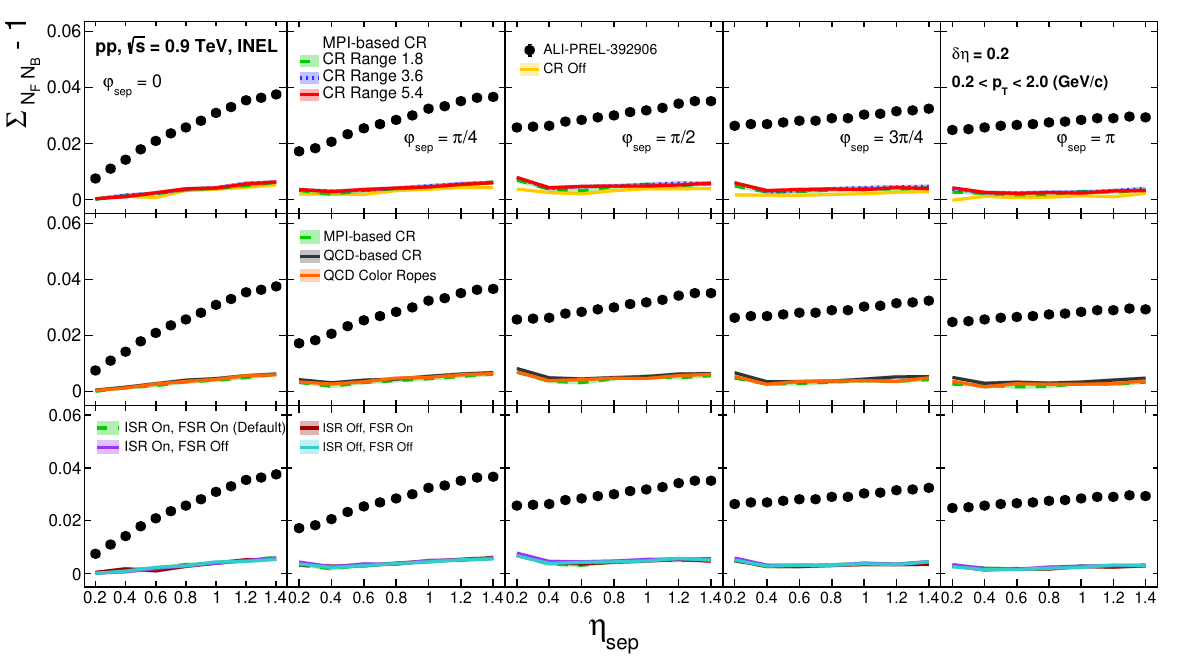}
    \caption{(Color online) Strongly intensive quantity minus one, $\Sigma_{\rm N_{\rm F} \rm N_{\rm B}} -1$, for azimuthally separated $\eta$-$\varphi$ windows of $\varphi_{\rm sep}$ = 0, $\pi$/4, $\pi$/2, 3$\pi$/4 and $\pi$, plotted as a function of $\eta_{\rm sep}$ for \texttt{PYTHIA8} simulated INEL pp collisions at $\sqrt{s}$ = 0.9 TeV and compared with ALICE Preliminary results~\cite{erokhin2021}.}
    \label{fig:sigma_etasep_azimuthal}
\end{figure*}

\subsection{Strongly Intensive Quantity: Roles of CR Ranges, CR Mechanisms and Parton Showers}

\subsubsection{$\eta_{\rm sep}$ and $\sqrt{s}$ dependence of $\Sigma$}

Fig.~\ref{fig:siq_all} shows the strongly intensive quantity  minus one,
$\Sigma_{\rm N_{\rm F} \rm N_{\rm B}} - 1$, as a function of $\eta_{\rm sep}$ for two center‑of‑mass energies, $\sqrt{s}=0.9$ TeV and 7 TeV. We systematically varied the CR range and CR mechanism, as well as the parton‑shower configuration, to assess their effects. \texttt{PYTHIA8} predictions reproduced the overall $\eta_{\rm sep}$‑dependent
trend seen in ALICE preliminary data~\cite{erokhin2021}, but underestimated the magnitude at both energies. In the \textit{left column} of Fig.~\ref{fig:siq_all}, the quantity remained essentially independent of CR range at 0.9 TeV, whereas at 7 TeV it increased slightly with CR range; the CR Off setting yielded the lowest values.  At $\sqrt{s} = 0.9$ TeV, the quantity $\Sigma_{\rm N_{\rm F} \rm N_{\rm B}} - 1$ showed no dependency on the CR range, as evident from the \textit{left column} and its inset.
\par
In the \textit{mid-column} of Fig.~\ref{fig:siq_all}, we plot the strongly intensive observable
$\Sigma_{\rm N_{\rm F} \rm N_{\rm B}} - 1$ as a function of $\eta_{\rm sep}$ for three CR mechanisms: MPI‑based (default), QCD‑based, and QCD Color Rope. Across the full $\eta_{\rm sep}$ range, the MPI‑based CR consistently yields the lowest values of $\Sigma_{N_{\rm F}N_{\rm B}} - 1$, whereas the QCD-based and Color Rope models produce nearly overlapping curves, a feature clearly supported by the inset.
\par
In the \textit{right column} of Fig.~\ref{fig:siq_all}, we plot
$\Sigma_{\rm N_{\rm F} \rm N_{\rm B}} - 1$ as a function of $\eta_{\rm sep}$ for the four possible combinations of initial‑state radiation (ISR) and final‑state radiation (FSR): ISR On/FSR On, ISR On/FSR Off, ISR Off/FSR On, and ISR Off/FSR Off. The curves corresponding to ISR On/FSR On (the default setting) and ISR Off/FSR On lie almost exactly on top of each other throughout the full $\eta_{\rm sep}$ range, implying that turning off ISR has a negligible effect when FSR remains active. Similarly, the ISR On/FSR Off and ISR Off/FSR Off curves also coincide, confirming that ISR in isolation (i.e., with FSR turned off) has no significant impact on the observable. 


\subsubsection{$\eta-\varphi$ dependence of $\Sigma$}

Finally, Fig.~\ref{fig:sigma_etasep_azimuthal} presents the dependence of the strongly intensive observable $\Sigma_{\rm N_{\rm F} \rm N_{\rm B}} - 1$ on $\eta_{\rm sep}$ across five distinct azimuthal sectors for pp collisions at $\sqrt{s}$ = 0.9 TeV, alongside preliminary results from ALICE~\cite{erokhin2021}. Computations employing the \texttt{PYTHIA8} event generator demonstrate qualitative agreement with the experimental data. Specifically, the model somewhat replicates the characteristic rise in $\Sigma_{\rm N_{\rm F} \rm N_{\rm B}} - 1$ within the narrow phase space region ($\varphi_{\rm sep} \leq \pi/4$) and its subsequent saturation for wider azimuthal separations ($\varphi_{\rm sep} \geq \pi/2$). To assess model sensitivity, three distinct CR ranges, underlying CR mechanisms, and parton shower implementations within \texttt{PYTHIA8} were investigated. While capturing the qualitative trend, none achieve quantitative agreement with the ALICE measurements across the full $\eta_{\rm sep}$ and $\varphi_{\rm sep}$ domain. A notable finding is the minimal variation observed in $\Sigma_{\rm N_{\rm F} \rm N_{\rm B}} - 1$ across all tested \texttt{PYTHIA8} configurations. This robustness is consistent with the fundamental property of strongly intensive quantities: their inherent insensitivity to volume (or participant number) fluctuations. Complementary studies, illustrated in Fig.~\ref{fig:bcorr_nsep}, corroborate the role of correlation dynamics, i.e., SRCs dominate the particle production mechanisms within narrow azimuthal acceptance, while LRCs become progressively more influential as the phase space widens. These findings underscore that string fragmentation models, as implemented in \texttt{PYTHIA8}, capture essential qualitative features of the correlation structure. To bridge this quantitative gap, the model must incorporate more refined physics ingredients: beyond adjusted color‑reconnection dynamics (such as string shoving or elevated rope tension).


\section{Summary}\label{section6}

In this investigation, we have presented a comprehensive study that underscores the influence of non-perturbative aspects of QCD in terms of MPI, CR, and QCD parton showers on particle production, analyzed through extensive, intensive, and strongly intensive Forward-Backward (FB) correlation measures in pp collisions deploying the \texttt{PYTHIA8} framework (version 8.311) at LHC energies, specifically at $\sqrt{s}$ = 0.9, 2.76 and 7 TeV. FB correlations were analyzed in terms of charged-particle multiplicity, summed transverse momentum and mean transverse momentum, and studied as a function of $\eta_{\rm gap}$ and $\eta_{\rm sep}$ using different variants of the \texttt{PYTHIA8} model and compared with the available experimental data~\cite{aad2012atlas,adam2015}. While the strongly intensive quantity ($\Sigma_{\rm N_{\rm F} \rm N_{\rm B}}$) grows with increasing $\eta_{\rm sep}$, the correlation strength for the extensive variables ($b_{\rm corr}(mult.),~b_{\rm corr}^{\sum p_{\rm T}}$) and for the intensive variable ($b_{\rm corr}^{\overline p_{\rm T}}$) decrease with increasing $\eta_{\rm gap}$. This behavior can be attributed to the fact that short-range correlations remain significant only within a limited pseudorapidity interval. The increase of $\Sigma_{\rm N_{\rm f}, N_{\rm b}}$ with $\eta_{\rm sep}$ signifies the existence of LRCs at larger pseudorapidity separations. Further, \textit{extensive} observables tend to decrease with increasing CR range. Among the CR strength tested, $b_{\rm corr}(mult.)$ appears to align well with the ALICE~\cite{adam2015} data at CR ranges 3.6 and 5.4 (of MPI-based CR) along with Color Ropes, offering a better description compared to CR range 1.8, across all energies. However, none of the ranges could explain $b_{\rm corr}^{\sum p_{\rm T}}$ published results from ATLAS~\cite{aad2012atlas}. Similarly, the above observables were studied in separated azimuth and pseudorapidity, where it was found that the azimuthal sectors with $\varphi_{\rm sep}>\pi/4$ were predominantly affected by LRCs, whereas those with $\varphi_{\rm sep} \leq \pi/4$ were primarily driven by SRCs. The specific interplay of MPI and CR is elucidated in the subsequent section, indicating that MPI and CR are crucial in shaping correlations. Further, parton shower dependence was examined across all the observables, which revealed some critical findings. It turned out that the magnitude of the extensive quantities ($b_{\rm corr}(mult.),~b_{\rm corr}^{\sum p_{\rm T}}$) decreased as ISR and/or FSR was turned off. This arises because suppressed shower radiation reduces secondary particle production~\cite{bierlich2022}, directly diminishing volume-dependent correlations in extensive observables. Conversely, for intensive observables, QCD radiations emerged as the source of SRCs, however, for strongly intensive quantity, FSR is a dominant driver. 


\nocite{*}

\begin{thebibliography}{83}%
\makeatletter
\providecommand \@ifxundefined [1]{%
 \@ifx{#1\undefined}
}%
\providecommand \@ifnum [1]{%
 \ifnum #1\expandafter \@firstoftwo
 \else \expandafter \@secondoftwo
 \fi
}%
\providecommand \@ifx [1]{%
 \ifx #1\expandafter \@firstoftwo
 \else \expandafter \@secondoftwo
 \fi
}%
\providecommand \natexlab [1]{#1}%
\providecommand \enquote  [1]{``#1''}%
\providecommand \bibnamefont  [1]{#1}%
\providecommand \bibfnamefont [1]{#1}%
\providecommand \citenamefont [1]{#1}%
\providecommand \href@noop [0]{\@secondoftwo}%
\providecommand \href [0]{\begingroup \@sanitize@url \@href}%
\providecommand \@href[1]{\@@startlink{#1}\@@href}%
\providecommand \@@href[1]{\endgroup#1\@@endlink}%
\providecommand \@sanitize@url [0]{\catcode `\\12\catcode `\$12\catcode `\&12\catcode `\#12\catcode `\^12\catcode `\_12\catcode `\%12\relax}%
\providecommand \@@startlink[1]{}%
\providecommand \@@endlink[0]{}%
\providecommand \url  [0]{\begingroup\@sanitize@url \@url }%
\providecommand \@url [1]{\endgroup\@href {#1}{\urlprefix }}%
\providecommand \urlprefix  [0]{URL }%
\providecommand \Eprint [0]{\href }%
\providecommand \doibase [0]{https://doi.org/}%
\providecommand \selectlanguage [0]{\@gobble}%
\providecommand \bibinfo  [0]{\@secondoftwo}%
\providecommand \bibfield  [0]{\@secondoftwo}%
\providecommand \translation [1]{[#1]}%
\providecommand \BibitemOpen [0]{}%
\providecommand \bibitemStop [0]{}%
\providecommand \bibitemNoStop [0]{.\EOS\space}%
\providecommand \EOS [0]{\spacefactor3000\relax}%
\providecommand \BibitemShut  [1]{\csname bibitem#1\endcsname}%
\let\auto@bib@innerbib\@empty
\bibitem [{\citenamefont {Shuryak}(1980)}]{shuryak1980}%
  \BibitemOpen
  \bibfield  {author} {\bibinfo {author} {\bibfnamefont {E.~V.}\ \bibnamefont {Shuryak}},\ }\href {https://doi.org/10.1016/0370-1573(80)90105-2} {\bibfield  {journal} {\bibinfo  {journal} {Phys. Rept.}\ }\textbf {\bibinfo {volume} {61}},\ \bibinfo {pages} {71} (\bibinfo {year} {1980})}\BibitemShut {NoStop}%
\bibitem [{\citenamefont {McLerran}(1986)}]{mclarren1986}%
  \BibitemOpen
  \bibfield  {author} {\bibinfo {author} {\bibfnamefont {L.}~\bibnamefont {McLerran}},\ }\href {https://doi.org/10.1103/RevModPhys.58.1021} {\bibfield  {journal} {\bibinfo  {journal} {Rev. Mod. Phys.}\ }\textbf {\bibinfo {volume} {58}},\ \bibinfo {pages} {1021} (\bibinfo {year} {1986})}\BibitemShut {NoStop}%
\bibitem [{\citenamefont {B.~Back}\ \emph {et~al.}(2005)\citenamefont {B.~Back} \emph {et~al.}}]{bbback2005phobos}%
  \BibitemOpen
  \bibfield  {author} {\bibinfo {author} {\bibfnamefont {B.}~\bibnamefont {B.~Back}} \emph {et~al.} (\bibinfo {collaboration} {PHOBOS Collaboration}),\ }\href {https://doi.org/10.1016/j.nuclphysa.2005.03.084} {\bibfield  {journal} {\bibinfo  {journal} {Nucl. Phys. A}\ }\textbf {\bibinfo {volume} {757}},\ \bibinfo {pages} {28} (\bibinfo {year} {2005})}\BibitemShut {NoStop}%
\bibitem [{\citenamefont {Adams}\ \emph {et~al.}(2005)\citenamefont {Adams} \emph {et~al.}}]{adams2005star}%
  \BibitemOpen
  \bibfield  {author} {\bibinfo {author} {\bibfnamefont {J.}~\bibnamefont {Adams}} \emph {et~al.} (\bibinfo {collaboration} {STAR Collaboration}),\ }\href {https://doi.org/10.1016/j.nuclphysa.2005.03.085} {\bibfield  {journal} {\bibinfo  {journal} {Nucl. Phys. A}\ }\textbf {\bibinfo {volume} {757}},\ \bibinfo {pages} {102} (\bibinfo {year} {2005})}\BibitemShut {NoStop}%
\bibitem [{\citenamefont {Adcox}\ \emph {et~al.}(2005)\citenamefont {Adcox} \emph {et~al.}}]{adcox2005phenix}%
  \BibitemOpen
  \bibfield  {author} {\bibinfo {author} {\bibfnamefont {K.}~\bibnamefont {Adcox}} \emph {et~al.} (\bibinfo {collaboration} {PHENIX Collaboration}),\ }\href {https://doi.org/10.1016/j.nuclphysa.2005.03.086} {\bibfield  {journal} {\bibinfo  {journal} {Nucl. Phys. A}\ }\textbf {\bibinfo {volume} {757}},\ \bibinfo {pages} {184} (\bibinfo {year} {2005})}\BibitemShut {NoStop}%
\bibitem [{\citenamefont {Alner}\ \emph {et~al.}(1987)\citenamefont {Alner} \emph {et~al.}}]{ua5alner1987}%
  \BibitemOpen
  \bibfield  {author} {\bibinfo {author} {\bibfnamefont {G.~J.}\ \bibnamefont {Alner}} \emph {et~al.} (\bibinfo {collaboration} {UA5 Collaboration}),\ }\href {https://doi.org/10.1016/0370-1573(87)90130-X} {\bibfield  {journal} {\bibinfo  {journal} {Phys. Rep.}\ }\textbf {\bibinfo {volume} {154}},\ \bibinfo {pages} {247} (\bibinfo {year} {1987})}\BibitemShut {NoStop}%
\bibitem [{\citenamefont {Uhlig}\ \emph {et~al.}(1978)\citenamefont {Uhlig} \emph {et~al.}}]{uhlig1978}%
  \BibitemOpen
  \bibfield  {author} {\bibinfo {author} {\bibfnamefont {S.}~\bibnamefont {Uhlig}} \emph {et~al.},\ }\href {https://doi.org/10.1016/0550-3213(78)90254-7} {\bibfield  {journal} {\bibinfo  {journal} {Nucl. Phys. B}\ }\textbf {\bibinfo {volume} {132}},\ \bibinfo {pages} {15} (\bibinfo {year} {1978})}\BibitemShut {NoStop}%
\bibitem [{\citenamefont {Werner}(2008)}]{werner2008}%
  \BibitemOpen
  \bibfield  {author} {\bibinfo {author} {\bibfnamefont {K.}~\bibnamefont {Werner}},\ }\href {https://doi.org/10.1016/j.nuclphysbps.2007.10.012} {\bibfield  {journal} {\bibinfo  {journal} {Nucl. Phys. B (Proc. Suppl.)}\ }\textbf {\bibinfo {volume} {175}},\ \bibinfo {pages} {81} (\bibinfo {year} {2008})}\BibitemShut {NoStop}%
\bibitem [{\citenamefont {Werner}\ \emph {et~al.}(2014)\citenamefont {Werner} \emph {et~al.}}]{werner2014}%
  \BibitemOpen
  \bibfield  {author} {\bibinfo {author} {\bibfnamefont {K.}~\bibnamefont {Werner}} \emph {et~al.},\ }\href {https://doi.org/10.1103/PhysRevC.89.064903} {\bibfield  {journal} {\bibinfo  {journal} {Phys. Rev. C}\ }\textbf {\bibinfo {volume} {89}},\ \bibinfo {pages} {064903} (\bibinfo {year} {2014})}\BibitemShut {NoStop}%
\bibitem [{\citenamefont {V.~Bravina}\ \emph {et~al.}(2018)\citenamefont {V.~Bravina} \emph {et~al.}}]{bravina2018}%
  \BibitemOpen
  \bibfield  {author} {\bibinfo {author} {\bibfnamefont {L.}~\bibnamefont {V.~Bravina}} \emph {et~al.},\ }\href {https://doi.org/10.1016/j.physletb.2018.10.053} {\bibfield  {journal} {\bibinfo  {journal} {Phys. Lett. B}\ }\textbf {\bibinfo {volume} {787}},\ \bibinfo {pages} {146} (\bibinfo {year} {2018})}\BibitemShut {NoStop}%
\bibitem [{\citenamefont {Capella}\ \emph {et~al.}(1994)\citenamefont {Capella} \emph {et~al.}}]{capella1994}%
  \BibitemOpen
  \bibfield  {author} {\bibinfo {author} {\bibfnamefont {A.}~\bibnamefont {Capella}} \emph {et~al.},\ }\href {https://doi.org/10.1016/0370-1573(94)90064-7} {\bibfield  {journal} {\bibinfo  {journal} {Phys. Rept.}\ }\textbf {\bibinfo {volume} {236}},\ \bibinfo {pages} {225} (\bibinfo {year} {1994})}\BibitemShut {NoStop}%
\bibitem [{\citenamefont {Capella}\ and\ \citenamefont {Krzywicki}(1978)}]{capella1978}%
  \BibitemOpen
  \bibfield  {author} {\bibinfo {author} {\bibfnamefont {A.}~\bibnamefont {Capella}}\ and\ \bibinfo {author} {\bibfnamefont {A.}~\bibnamefont {Krzywicki}},\ }\href {https://doi.org/10.1103/PhysRevD.18.4120} {\bibfield  {journal} {\bibinfo  {journal} {Phys. Rev. D}\ }\textbf {\bibinfo {volume} {18}},\ \bibinfo {pages} {4120} (\bibinfo {year} {1978})}\BibitemShut {NoStop}%
\bibitem [{\citenamefont {Vechernin}\ and\ \citenamefont {Kolevatov}(2007)}]{vechernin2007multiplicity}%
  \BibitemOpen
  \bibfield  {author} {\bibinfo {author} {\bibfnamefont {V.}~\bibnamefont {Vechernin}}\ and\ \bibinfo {author} {\bibfnamefont {R.}~\bibnamefont {Kolevatov}},\ }\href {https://doi.org/10.1134/S1063778807100158} {\bibfield  {journal} {\bibinfo  {journal} {Phys. Atom. Nuclei}\ }\textbf {\bibinfo {volume} {70}},\ \bibinfo {pages} {1797} (\bibinfo {year} {2007})}\BibitemShut {NoStop}%
\bibitem [{\citenamefont {Braun}\ \emph {et~al.}(2004)\citenamefont {Braun} \emph {et~al.}}]{braun2004}%
  \BibitemOpen
  \bibfield  {author} {\bibinfo {author} {\bibfnamefont {M.~A.}\ \bibnamefont {Braun}} \emph {et~al.},\ }\href {https://doi.org/10.1140/epjc/s2003-01443-6} {\bibfield  {journal} {\bibinfo  {journal} {Eur. Phys. J. C}\ }\textbf {\bibinfo {volume} {32}},\ \bibinfo {pages} {535} (\bibinfo {year} {2004})}\BibitemShut {NoStop}%
\bibitem [{\citenamefont {Braun}\ and\ \citenamefont {Pajares}(2000)}]{braun2000implications}%
  \BibitemOpen
  \bibfield  {author} {\bibinfo {author} {\bibfnamefont {M.}~\bibnamefont {Braun}}\ and\ \bibinfo {author} {\bibfnamefont {C.}~\bibnamefont {Pajares}},\ }\href {https://doi.org/10.1007/s100520050027} {\bibfield  {journal} {\bibinfo  {journal} {Eur. Phys. J. C.}\ }\textbf {\bibinfo {volume} {16}},\ \bibinfo {pages} {349} (\bibinfo {year} {2000})}\BibitemShut {NoStop}%
\bibitem [{\citenamefont {Gorenstein}\ and\ \citenamefont {Ga{\'z}dzicki}(2011)}]{gorenstein2011}%
  \BibitemOpen
  \bibfield  {author} {\bibinfo {author} {\bibfnamefont {M.~I.}\ \bibnamefont {Gorenstein}}\ and\ \bibinfo {author} {\bibfnamefont {M.}~\bibnamefont {Ga{\'z}dzicki}},\ }\href {https://doi.org/10.1103/PhysRevC.84.014904} {\bibfield  {journal} {\bibinfo  {journal} {Phys. Rev. C—Nucl. Phys.}\ }\textbf {\bibinfo {volume} {84}},\ \bibinfo {pages} {014904} (\bibinfo {year} {2011})}\BibitemShut {NoStop}%
\bibitem [{\citenamefont {Anticic}\ \emph {et~al.}(2015)\citenamefont {Anticic} \emph {et~al.}}]{anticic2015}%
  \BibitemOpen
  \bibfield  {author} {\bibinfo {author} {\bibfnamefont {T.}~\bibnamefont {Anticic}} \emph {et~al.} (\bibinfo {collaboration} {NA49 Collaboration}),\ }\href {https://doi.org/10.1103/PhysRevC.92.044905} {\bibfield  {journal} {\bibinfo  {journal} {Phys. Rev. C}\ }\textbf {\bibinfo {volume} {92}},\ \bibinfo {pages} {044905} (\bibinfo {year} {2015})}\BibitemShut {NoStop}%
\bibitem [{\citenamefont {G.~Albrow}\ \emph {et~al.}(1978)\citenamefont {G.~Albrow} \emph {et~al.}}]{albrow1978}%
  \BibitemOpen
  \bibfield  {author} {\bibinfo {author} {\bibfnamefont {M.}~\bibnamefont {G.~Albrow}} \emph {et~al.} (\bibinfo {collaboration} {British-French-Scandinavian Collaboration}),\ }\href {https://doi.org/10.1016/0550-3213(78)90088-3} {\bibfield  {journal} {\bibinfo  {journal} {Nucl. Phys. B}\ }\textbf {\bibinfo {volume} {145}},\ \bibinfo {pages} {305} (\bibinfo {year} {1978})}\BibitemShut {NoStop}%
\bibitem [{\citenamefont {Alpgard}\ \emph {et~al.}(1983)\citenamefont {Alpgard} \emph {et~al.}}]{alpgard1983}%
  \BibitemOpen
  \bibfield  {author} {\bibinfo {author} {\bibfnamefont {K.}~\bibnamefont {Alpgard}} \emph {et~al.} (\bibinfo {collaboration} {UA5 Collaboration}),\ }\href {https://doi.org/10.1016/0370-2693(83)91218-2} {\bibfield  {journal} {\bibinfo  {journal} {Phys. Lett. B}\ }\textbf {\bibinfo {volume} {123}},\ \bibinfo {pages} {361} (\bibinfo {year} {1983})}\BibitemShut {NoStop}%
\bibitem [{\citenamefont {Ansorge}\ \emph {et~al.}(1988)\citenamefont {Ansorge} \emph {et~al.}}]{ansorge1988}%
  \BibitemOpen
  \bibfield  {author} {\bibinfo {author} {\bibfnamefont {R.}~\bibnamefont {Ansorge}} \emph {et~al.} (\bibinfo {collaboration} {UA5 Collaboration}),\ }\href {https://doi.org/10.1007/BF01579906} {\bibfield  {journal} {\bibinfo  {journal} {Zeitschrift f{\"u}r Physik C Particles and Fields}\ }\textbf {\bibinfo {volume} {37}},\ \bibinfo {pages} {191} (\bibinfo {year} {1988})}\BibitemShut {NoStop}%
\bibitem [{\citenamefont {I.~Abelev}\ \emph {et~al.}(2009)\citenamefont {I.~Abelev} \emph {et~al.}}]{abelev2009}%
  \BibitemOpen
  \bibfield  {author} {\bibinfo {author} {\bibfnamefont {B.}~\bibnamefont {I.~Abelev}} \emph {et~al.} (\bibinfo {collaboration} {STAR Collaboration}),\ }\href {https://doi.org/10.1103/PhysRevLett.103.172301} {\bibfield  {journal} {\bibinfo  {journal} {Phys. Rev. Lett.}\ }\textbf {\bibinfo {volume} {103}},\ \bibinfo {pages} {172301} (\bibinfo {year} {2009})}\BibitemShut {NoStop}%
\bibitem [{\citenamefont {Srivastava}\ \emph {et~al.}(2007)\citenamefont {Srivastava} \emph {et~al.}}]{bksrivastava2007}%
  \BibitemOpen
  \bibfield  {author} {\bibinfo {author} {\bibfnamefont {B.~K.}\ \bibnamefont {Srivastava}} \emph {et~al.} (\bibinfo {collaboration} {STAR Collaboration}),\ }\href {https://doi.org/10.1142/S0218301307007702} {\bibfield  {journal} {\bibinfo  {journal} {Int. J. Mod. Phys. E}\ }\textbf {\bibinfo {volume} {16}},\ \bibinfo {pages} {2210} (\bibinfo {year} {2007})}\BibitemShut {NoStop}%
\bibitem [{\citenamefont {Srivastava}\ \emph {et~al.}(2008)\citenamefont {Srivastava} \emph {et~al.}}]{srivastava2008star2}%
  \BibitemOpen
  \bibfield  {author} {\bibinfo {author} {\bibfnamefont {B.~K.}\ \bibnamefont {Srivastava}} \emph {et~al.} (\bibinfo {collaboration} {STAR Collaboration}),\ }\href {https://dx.doi.org/10.1088/0954-3899/35/10/104140} {\bibfield  {journal} {\bibinfo  {journal} {J. Phys. G: Nucl. and Particle Phys.}\ }\textbf {\bibinfo {volume} {35}},\ \bibinfo {pages} {104140} (\bibinfo {year} {2008})}\BibitemShut {NoStop}%
\bibitem [{\citenamefont {Tarnowsky}\ \emph {et~al.}(2007)\citenamefont {Tarnowsky} \emph {et~al.}}]{tarnowsky2007}%
  \BibitemOpen
  \bibfield  {author} {\bibinfo {author} {\bibfnamefont {T.}~\bibnamefont {Tarnowsky}} \emph {et~al.} (\bibinfo {collaboration} {STAR Collaboration}),\ }\href {https://doi.org/10.1142/S0218301307009348} {\bibfield  {journal} {\bibinfo  {journal} {Int. J. Mod. Phys. E}\ }\textbf {\bibinfo {volume} {16}},\ \bibinfo {pages} {3363} (\bibinfo {year} {2007})}\BibitemShut {NoStop}%
\bibitem [{\citenamefont {Aad}\ \emph {et~al.}(2012)\citenamefont {Aad} \emph {et~al.}}]{aad2012atlas}%
  \BibitemOpen
  \bibfield  {author} {\bibinfo {author} {\bibfnamefont {G.}~\bibnamefont {Aad}} \emph {et~al.} (\bibinfo {collaboration} {ATLAS Collaboration}),\ }\href {https://doi.org/10.1007/JHEP07(2012)019} {\bibfield  {journal} {\bibinfo  {journal} {J. High Energy Phys.}\ }\textbf {\bibinfo {volume} {07}}\bibinfo  {number} { (2012)},\ \bibinfo {pages} {019}}\BibitemShut {NoStop}%
\bibitem [{\citenamefont {Adam}\ \emph {et~al.}(2015)\citenamefont {Adam} \emph {et~al.}}]{adam2015}%
  \BibitemOpen
\bibfield  {number} {  }\bibfield  {author} {\bibinfo {author} {\bibfnamefont {J.}~\bibnamefont {Adam}} \emph {et~al.} (\bibinfo {collaboration} {ALICE Collaboration}),\ }\href {https://doi.org/10.1007/JHEP05(2015)097} {\bibfield  {journal} {\bibinfo  {journal} {J. High Energy Phys.}\ }\textbf {\bibinfo {volume} {05}}\bibinfo  {number} { (2015)},\ \bibinfo {pages} {097}}\BibitemShut {NoStop}%
\bibitem [{\citenamefont {Altsybeev}(2017)}]{altsybeev2017}%
  \BibitemOpen
\bibfield  {number} {  }\bibfield  {author} {\bibinfo {author} {\bibfnamefont {I.}~\bibnamefont {Altsybeev}} (\bibinfo {collaboration} {ALICE Collaboration}),\ }\href {https://arxiv.org/abs/1711.04844v1} {\bibfield  {journal} {\bibinfo  {journal} {arXiv:1711.04844}\ } (\bibinfo {year} {2017})}\BibitemShut {NoStop}%
\bibitem [{\citenamefont {Armesto}\ \emph {et~al.}(2007)\citenamefont {Armesto}, \citenamefont {McLerran},\ and\ \citenamefont {Pajares}}]{armesto2007}%
  \BibitemOpen
  \bibfield  {author} {\bibinfo {author} {\bibfnamefont {N.}~\bibnamefont {Armesto}}, \bibinfo {author} {\bibfnamefont {L.}~\bibnamefont {McLerran}},\ and\ \bibinfo {author} {\bibfnamefont {C.}~\bibnamefont {Pajares}},\ }\href {https://doi.org/10.1016/j.nuclphysa.2006.10.074} {\bibfield  {journal} {\bibinfo  {journal} {Nucl. Phys. A}\ }\textbf {\bibinfo {volume} {781}},\ \bibinfo {pages} {201} (\bibinfo {year} {2007})}\BibitemShut {NoStop}%
\bibitem [{\citenamefont {P.~Konchakovski}\ \emph {et~al.}(2009)\citenamefont {P.~Konchakovski} \emph {et~al.}}]{konchakovski2009}%
  \BibitemOpen
  \bibfield  {author} {\bibinfo {author} {\bibfnamefont {V.}~\bibnamefont {P.~Konchakovski}} \emph {et~al.},\ }\href {https://doi.org/10.1103/PhysRevC.79.034910} {\bibfield  {journal} {\bibinfo  {journal} {Phys. Rev. C}\ }\textbf {\bibinfo {volume} {79}},\ \bibinfo {pages} {034910} (\bibinfo {year} {2009})}\BibitemShut {NoStop}%
\bibitem [{\citenamefont {Brogueira}\ \emph {et~al.}(2009)\citenamefont {Brogueira}, \citenamefont {de~Deus},\ and\ \citenamefont {Pajares}}]{brogueira2009}%
  \BibitemOpen
  \bibfield  {author} {\bibinfo {author} {\bibfnamefont {P.}~\bibnamefont {Brogueira}}, \bibinfo {author} {\bibfnamefont {J.~D.}\ \bibnamefont {de~Deus}},\ and\ \bibinfo {author} {\bibfnamefont {C.}~\bibnamefont {Pajares}},\ }\href {https://doi.org/10.1016/j.physletb.2009.04.025} {\bibfield  {journal} {\bibinfo  {journal} {Phys. Lett. B}\ }\textbf {\bibinfo {volume} {675}},\ \bibinfo {pages} {308} (\bibinfo {year} {2009})}\BibitemShut {NoStop}%
\bibitem [{\citenamefont {Bzdak}(2009)}]{bzdak2009}%
  \BibitemOpen
  \bibfield  {author} {\bibinfo {author} {\bibfnamefont {A.}~\bibnamefont {Bzdak}},\ }\href {https://doi.org/10.1103/PhysRevC.80.024906} {\bibfield  {journal} {\bibinfo  {journal} {Phys. Rev. C}\ }\textbf {\bibinfo {volume} {80}},\ \bibinfo {pages} {024906} (\bibinfo {year} {2009})}\BibitemShut {NoStop}%
\bibitem [{\citenamefont {Yan}\ \emph {et~al.}(2009)\citenamefont {Yan} \emph {et~al.}}]{yan2009}%
  \BibitemOpen
  \bibfield  {author} {\bibinfo {author} {\bibfnamefont {Y.-L.}\ \bibnamefont {Yan}} \emph {et~al.},\ }\href {https://doi.org/10.1103/PhysRevC.79.054902} {\bibfield  {journal} {\bibinfo  {journal} {Phys. Rev. C}\ }\textbf {\bibinfo {volume} {79}},\ \bibinfo {pages} {054902} (\bibinfo {year} {2009})}\BibitemShut {NoStop}%
\bibitem [{\citenamefont {Yan}\ \emph {et~al.}(2010)\citenamefont {Yan} \emph {et~al.}}]{yan2010}%
  \BibitemOpen
  \bibfield  {author} {\bibinfo {author} {\bibfnamefont {Y.-L.}\ \bibnamefont {Yan}} \emph {et~al.},\ }\href {https://doi.org/10.1103/PhysRevC.81.044914} {\bibfield  {journal} {\bibinfo  {journal} {Phys. Rev. C}\ }\textbf {\bibinfo {volume} {81}},\ \bibinfo {pages} {044914} (\bibinfo {year} {2010})}\BibitemShut {NoStop}%
\bibitem [{\citenamefont {Lappi}\ and\ \citenamefont {McLerran}(2010)}]{lappi2010}%
  \BibitemOpen
  \bibfield  {author} {\bibinfo {author} {\bibfnamefont {T.}~\bibnamefont {Lappi}}\ and\ \bibinfo {author} {\bibfnamefont {L.}~\bibnamefont {McLerran}},\ }\href {https://doi.org/10.1016/j.nuclphysa.2009.11.003} {\bibfield  {journal} {\bibinfo  {journal} {Nucl. Phys. A}\ }\textbf {\bibinfo {volume} {832}},\ \bibinfo {pages} {330} (\bibinfo {year} {2010})}\BibitemShut {NoStop}%
\bibitem [{\citenamefont {Bialas}\ and\ \citenamefont {Zalewski}(2011)}]{bialas2011}%
  \BibitemOpen
  \bibfield  {author} {\bibinfo {author} {\bibfnamefont {A.}~\bibnamefont {Bialas}}\ and\ \bibinfo {author} {\bibfnamefont {K.}~\bibnamefont {Zalewski}},\ }\href {https://doi.org/10.1016/j.physletb.2011.03.036} {\bibfield  {journal} {\bibinfo  {journal} {Phys. Lett. B}\ }\textbf {\bibinfo {volume} {698}},\ \bibinfo {pages} {416} (\bibinfo {year} {2011})}\BibitemShut {NoStop}%
\bibitem [{\citenamefont {Mondal}\ \emph {et~al.}(2020)\citenamefont {Mondal} \emph {et~al.}}]{mondal2020}%
  \BibitemOpen
  \bibfield  {author} {\bibinfo {author} {\bibfnamefont {M.}~\bibnamefont {Mondal}} \emph {et~al.},\ }\href {https://doi.org/10.1103/PhysRevD.102.014033} {\bibfield  {journal} {\bibinfo  {journal} {Phys. Rev. D}\ }\textbf {\bibinfo {volume} {102}},\ \bibinfo {pages} {014033} (\bibinfo {year} {2020})}\BibitemShut {NoStop}%
\bibitem [{\citenamefont {Mondal}\ \emph {et~al.}(2023)\citenamefont {Mondal} \emph {et~al.}}]{mondal2023}%
  \BibitemOpen
  \bibfield  {author} {\bibinfo {author} {\bibfnamefont {J.}~\bibnamefont {Mondal}} \emph {et~al.},\ }\href {https://doi.org/10.1103/PhysRevD.107.114016} {\bibfield  {journal} {\bibinfo  {journal} {Phys. Rev. D}\ }\textbf {\bibinfo {volume} {107}},\ \bibinfo {pages} {114016} (\bibinfo {year} {2023})}\BibitemShut {NoStop}%
\bibitem [{\citenamefont {B.~Kaidalov}(2003)}]{kaidalov2003}%
  \BibitemOpen
  \bibfield  {author} {\bibinfo {author} {\bibfnamefont {A.}~\bibnamefont {B.~Kaidalov}},\ }\href {https://doi.org/10.1134/1.1625743} {\bibfield  {journal} {\bibinfo  {journal} {Phys. Atom. Nuclei}\ }\textbf {\bibinfo {volume} {66}},\ \bibinfo {pages} {1994} (\bibinfo {year} {2003})}\BibitemShut {NoStop}%
\bibitem [{\citenamefont {S.~Amelin}\ \emph {et~al.}(1993)\citenamefont {S.~Amelin} \emph {et~al.}}]{amelin1993plb}%
  \BibitemOpen
  \bibfield  {author} {\bibinfo {author} {\bibfnamefont {N.}~\bibnamefont {S.~Amelin}} \emph {et~al.},\ }\href {https://doi.org/10.1016/0370-2693(93)90085-V} {\bibfield  {journal} {\bibinfo  {journal} {Phys. Lett. B}\ }\textbf {\bibinfo {volume} {306}},\ \bibinfo {pages} {312} (\bibinfo {year} {1993})}\BibitemShut {NoStop}%
\bibitem [{\citenamefont {S.~Amelin}\ \emph {et~al.}(1994{\natexlab{a}})\citenamefont {S.~Amelin} \emph {et~al.}}]{amelin1994prl}%
  \BibitemOpen
  \bibfield  {author} {\bibinfo {author} {\bibfnamefont {N.}~\bibnamefont {S.~Amelin}} \emph {et~al.},\ }\href {https://doi.org/10.1103/PhysRevLett.73.2813} {\bibfield  {journal} {\bibinfo  {journal} {Phys. Rev. Lett.}\ }\textbf {\bibinfo {volume} {73}},\ \bibinfo {pages} {2813} (\bibinfo {year} {1994}{\natexlab{a}})}\BibitemShut {NoStop}%
\bibitem [{\citenamefont {S.~Amelin}\ \emph {et~al.}(1994{\natexlab{b}})\citenamefont {S.~Amelin} \emph {et~al.}}]{amelin1994zpc}%
  \BibitemOpen
  \bibfield  {author} {\bibinfo {author} {\bibfnamefont {N.}~\bibnamefont {S.~Amelin}} \emph {et~al.},\ }\href {https://doi.org/10.1007/BF01580331} {\bibfield  {journal} {\bibinfo  {journal} {Z. Phys. C- Part. and Fields}\ }\textbf {\bibinfo {volume} {63}},\ \bibinfo {pages} {507} (\bibinfo {year} {1994}{\natexlab{b}})}\BibitemShut {NoStop}%
\bibitem [{\citenamefont {Brogueira}\ \emph {et~al.}(2007)\citenamefont {Brogueira}, \citenamefont {de~Deus},\ and\ \citenamefont {Pajares}}]{brogueira2007}%
  \BibitemOpen
  \bibfield  {author} {\bibinfo {author} {\bibfnamefont {P.}~\bibnamefont {Brogueira}}, \bibinfo {author} {\bibfnamefont {J.~D.}\ \bibnamefont {de~Deus}},\ and\ \bibinfo {author} {\bibfnamefont {C.}~\bibnamefont {Pajares}},\ }\href {https://doi.org/10.1016/j.physletb.2009.04.025} {\bibfield  {journal} {\bibinfo  {journal} {Phys. Lett. B}\ }\textbf {\bibinfo {volume} {653}},\ \bibinfo {pages} {202} (\bibinfo {year} {2007})}\BibitemShut {NoStop}%
\bibitem [{\citenamefont {Khan}\ \emph {et~al.}(2019)\citenamefont {Khan} \emph {et~al.}}]{khan2019}%
  \BibitemOpen
  \bibfield  {author} {\bibinfo {author} {\bibfnamefont {R.}~\bibnamefont {Khan}} \emph {et~al.},\ }\href {https://dx.doi.org/10.1088/0253-6102/71/10/1172} {\bibfield  {journal} {\bibinfo  {journal} {Commun. Theor. Phys.}\ }\textbf {\bibinfo {volume} {71}},\ \bibinfo {pages} {1172} (\bibinfo {year} {2019})}\BibitemShut {NoStop}%
\bibitem [{\citenamefont {Bhattacharyya}(2024)}]{bhattacharyya2024}%
  \BibitemOpen
  \bibfield  {author} {\bibinfo {author} {\bibfnamefont {S.}~\bibnamefont {Bhattacharyya}},\ }\href {https://doi.org/10.1140/epjp/s13360-024-04918-5} {\bibfield  {journal} {\bibinfo  {journal} {Eur. Phys. J. Plus}\ }\textbf {\bibinfo {volume} {139}},\ \bibinfo {pages} {122} (\bibinfo {year} {2024})}\BibitemShut {NoStop}%
\bibitem [{\citenamefont {Kundu}\ \emph {et~al.}(2019)\citenamefont {Kundu}, \citenamefont {Mohanty},\ and\ \citenamefont {Mallick}}]{kundu2019}%
  \BibitemOpen
  \bibfield  {author} {\bibinfo {author} {\bibfnamefont {S.}~\bibnamefont {Kundu}}, \bibinfo {author} {\bibfnamefont {B.}~\bibnamefont {Mohanty}},\ and\ \bibinfo {author} {\bibfnamefont {D.}~\bibnamefont {Mallick}},\ }\href {https://doi.org/10.48550/arXiv.1912.05176} {\bibfield  {journal} {\bibinfo  {journal} {arXiv:1912.05176}\ } (\bibinfo {year} {2019})}\BibitemShut {NoStop}%
\bibitem [{\citenamefont {Cuautle}\ \emph {et~al.}(2019)\citenamefont {Cuautle}, \citenamefont {Dominguez},\ and\ \citenamefont {Maldonado}}]{cuautle2019}%
  \BibitemOpen
  \bibfield  {author} {\bibinfo {author} {\bibfnamefont {E.}~\bibnamefont {Cuautle}}, \bibinfo {author} {\bibfnamefont {E.}~\bibnamefont {Dominguez}},\ and\ \bibinfo {author} {\bibfnamefont {I.}~\bibnamefont {Maldonado}},\ }\href {https://doi.org/10.1140/epjc/s10052-019-7128-2} {\bibfield  {journal} {\bibinfo  {journal} {Eur. Phys. J. C}\ }\textbf {\bibinfo {volume} {79}},\ \bibinfo {pages} {1} (\bibinfo {year} {2019})}\BibitemShut {NoStop}%
\bibitem [{\citenamefont {Erokhin}(2021)}]{erokhin2021}%
  \BibitemOpen
  \bibfield  {author} {\bibinfo {author} {\bibfnamefont {A.}~\bibnamefont {Erokhin}} (\bibinfo {collaboration} {ALICE Collaboration}),\ }in\ \href {https://indico.cern.ch/event/854124/contributions/4134683/} {\emph {\bibinfo {booktitle} {Initial Stages (IS) Proc.}}}\ (\bibinfo {year} {2021})\BibitemShut {NoStop}%
\bibitem [{\citenamefont {Andronov}\ and\ \citenamefont {Vechernin}(2019)}]{andronov2019}%
  \BibitemOpen
  \bibfield  {author} {\bibinfo {author} {\bibfnamefont {E.}~\bibnamefont {Andronov}}\ and\ \bibinfo {author} {\bibfnamefont {V.}~\bibnamefont {Vechernin}},\ }\href {https://doi.org/10.1140/epja/i2019-12681-x} {\bibfield  {journal} {\bibinfo  {journal} {Eur. Phys. J. A}\ }\textbf {\bibinfo {volume} {55}},\ \bibinfo {pages} {14} (\bibinfo {year} {2019})}\BibitemShut {NoStop}%
\bibitem [{\citenamefont {Prokhorova}(2018)}]{prokhorova2018pseudorapidity}%
  \BibitemOpen
  \bibfield  {author} {\bibinfo {author} {\bibfnamefont {D.}~\bibnamefont {Prokhorova}} (\bibinfo {collaboration} {NA61/SHINE Collaboration}),\ }\href {https://doi.org/10.48550/arXiv.1801.06690} {\bibfield  {journal} {\bibinfo  {journal} {arXiv:1801.06690}\ } (\bibinfo {year} {2018})}\BibitemShut {NoStop}%
\bibitem [{\citenamefont {Malik}\ \emph {et~al.}(2024)\citenamefont {Malik} \emph {et~al.}}]{malik2024}%
  \BibitemOpen
  \bibfield  {author} {\bibinfo {author} {\bibfnamefont {N.}~\bibnamefont {Malik}} \emph {et~al.},\ }in\ \href {https://sympnp.org/proceedings/68/D33.pdf} {\emph {\bibinfo {booktitle} {Proceedings of the DAE Symp. on Nucl. Phys.}}},\ Vol.~\bibinfo {volume} {68}\ (\bibinfo {year} {2024})\ p.\ \bibinfo {pages} {889}\BibitemShut {NoStop}%
\bibitem [{\citenamefont {Singh}\ \emph {et~al.}(2024)\citenamefont {Singh} \emph {et~al.}}]{singh2024}%
  \BibitemOpen
  \bibfield  {author} {\bibinfo {author} {\bibfnamefont {S.}~\bibnamefont {Singh}} \emph {et~al.},\ }in\ \href {https://sympnp.org/proceedings/68/D11.pdf} {\emph {\bibinfo {booktitle} {Proceedings of the DAE Symp. on Nucl. Phys.}}},\ Vol.~\bibinfo {volume} {68}\ (\bibinfo {year} {2024})\ p.\ \bibinfo {pages} {845}\BibitemShut {NoStop}%
\bibitem [{\citenamefont {Sputowska}(2023)}]{sputowska2023forward}%
  \BibitemOpen
  \bibfield  {author} {\bibinfo {author} {\bibfnamefont {I.}~\bibnamefont {Sputowska}},\ }\href {https://doi.org/10.1103/PhysRevC.108.014903} {\bibfield  {journal} {\bibinfo  {journal} {Phys. Rev. C}\ }\textbf {\bibinfo {volume} {108}},\ \bibinfo {pages} {014903} (\bibinfo {year} {2023})}\BibitemShut {NoStop}%
\bibitem [{\citenamefont {Sputowska}(2019)}]{sputowska2019forward}%
  \BibitemOpen
  \bibfield  {author} {\bibinfo {author} {\bibfnamefont {I.}~\bibnamefont {Sputowska}},\ }in\ \href {https://doi.org/10.3390/proceedings2019010014} {\emph {\bibinfo {booktitle} {MDPI Proc.}}},\ Vol.~\bibinfo {volume} {10}\ (\bibinfo {year} {2019})\ p.~\bibinfo {pages} {14}\BibitemShut {NoStop}%
\bibitem [{\citenamefont {Sputowska}(2022)}]{sputowska2022forward}%
  \BibitemOpen
  \bibfield  {author} {\bibinfo {author} {\bibfnamefont {I.}~\bibnamefont {Sputowska}},\ }in\ \href {https://doi.org/10.1051/epjconf/202227405003} {\emph {\bibinfo {booktitle} {Proceedings of EPJ Web of Conferences (ConfXV)}}},\ Vol.\ \bibinfo {volume} {274}\ (\bibinfo {year} {2022})\ p.~\bibinfo {pages} {7}\BibitemShut {NoStop}%
\bibitem [{\citenamefont {Bierlich}\ \emph {et~al.}(2022)\citenamefont {Bierlich} \emph {et~al.}}]{bierlich2022}%
  \BibitemOpen
  \bibfield  {author} {\bibinfo {author} {\bibfnamefont {C.}~\bibnamefont {Bierlich}} \emph {et~al.},\ }\href {https://scipost.org/10.21468/SciPostPhysCodeb.8} {\bibfield  {journal} {\bibinfo  {journal} {SciPost Phys. Codebases}\ ,\ \bibinfo {pages} {008}} (\bibinfo {year} {2022})}\BibitemShut {NoStop}%
\bibitem [{\citenamefont {Skands}\ \emph {et~al.}(2014)\citenamefont {Skands}, \citenamefont {Carrazza},\ and\ \citenamefont {Rojo}}]{skands2014}%
  \BibitemOpen
  \bibfield  {author} {\bibinfo {author} {\bibfnamefont {P.}~\bibnamefont {Skands}}, \bibinfo {author} {\bibfnamefont {S.}~\bibnamefont {Carrazza}},\ and\ \bibinfo {author} {\bibfnamefont {J.}~\bibnamefont {Rojo}},\ }\href {https://doi.org/10.1140/epjc/s10052-014-3024-y} {\bibfield  {journal} {\bibinfo  {journal} {Eur. Phys. J. C}\ }\textbf {\bibinfo {volume} {74}},\ \bibinfo {pages} {3024} (\bibinfo {year} {2014})}\BibitemShut {NoStop}%
\bibitem [{\citenamefont {Sjöstrand}\ \emph {et~al.}(2015)\citenamefont {Sjöstrand} \emph {et~al.}}]{sjostrand2015}%
  \BibitemOpen
  \bibfield  {author} {\bibinfo {author} {\bibfnamefont {T.}~\bibnamefont {Sjöstrand}} \emph {et~al.},\ }\href {https://doi.org/10.1016/j.cpc.2015.01.024} {\bibfield  {journal} {\bibinfo  {journal} {Comp. Phys. Comm.}\ }\textbf {\bibinfo {volume} {191}},\ \bibinfo {pages} {159} (\bibinfo {year} {2015})}\BibitemShut {NoStop}%
\bibitem [{\citenamefont {Kar}\ \emph {et~al.}(2017)\citenamefont {Kar} \emph {et~al.}}]{kar2017}%
  \BibitemOpen
  \bibfield  {author} {\bibinfo {author} {\bibfnamefont {S.}~\bibnamefont {Kar}} \emph {et~al.},\ }\href {https://doi.org/10.1103/PhysRevD.95.014016} {\bibfield  {journal} {\bibinfo  {journal} {Phys. Rev. D}\ }\textbf {\bibinfo {volume} {95}},\ \bibinfo {pages} {014016} (\bibinfo {year} {2017})}\BibitemShut {NoStop}%
\bibitem [{\citenamefont {Sarma}\ \emph {et~al.}(2023)\citenamefont {Sarma} \emph {et~al.}}]{sarma2023}%
  \BibitemOpen
  \bibfield  {author} {\bibinfo {author} {\bibfnamefont {P.}~\bibnamefont {Sarma}} \emph {et~al.},\ }\href {https://doi.org/10.1140/epja/s10050-023-00989-7} {\bibfield  {journal} {\bibinfo  {journal} {Eur. Phys. J. A}\ }\textbf {\bibinfo {volume} {59}},\ \bibinfo {pages} {76} (\bibinfo {year} {2023})}\BibitemShut {NoStop}%
\bibitem [{\citenamefont {Bierlich}\ and\ \citenamefont {Christiansen}(2015)}]{bierlich2015}%
  \BibitemOpen
  \bibfield  {author} {\bibinfo {author} {\bibfnamefont {C.}~\bibnamefont {Bierlich}}\ and\ \bibinfo {author} {\bibfnamefont {J.~R.}\ \bibnamefont {Christiansen}},\ }\href {https://doi.org/10.1103/PhysRevD.92.094010} {\bibfield  {journal} {\bibinfo  {journal} {Phys. Rev. D}\ }\textbf {\bibinfo {volume} {92}},\ \bibinfo {pages} {094010} (\bibinfo {year} {2015})}\BibitemShut {NoStop}%
\bibitem [{\citenamefont {Ortiz~Vel{\'a}squez}\ \emph {et~al.}(2013)\citenamefont {Ortiz~Vel{\'a}squez} \emph {et~al.}}]{ortiz2013}%
  \BibitemOpen
  \bibfield  {author} {\bibinfo {author} {\bibfnamefont {A.}~\bibnamefont {Ortiz~Vel{\'a}squez}} \emph {et~al.},\ }\href {https://doi.org/10.1103/PhysRevLett.111.042001} {\bibfield  {journal} {\bibinfo  {journal} {Phys. Rev. Lett.}\ }\textbf {\bibinfo {volume} {111}},\ \bibinfo {pages} {042001} (\bibinfo {year} {2013})}\BibitemShut {NoStop}%
\bibitem [{\citenamefont {Dey}\ \emph {et~al.}(2022)\citenamefont {Dey}, \citenamefont {N.~Mishra},\ and\ \citenamefont {K.~Tripathy}}]{dey2022}%
  \BibitemOpen
  \bibfield  {author} {\bibinfo {author} {\bibfnamefont {K.}~\bibnamefont {Dey}}, \bibinfo {author} {\bibfnamefont {A.}~\bibnamefont {N.~Mishra}},\ and\ \bibinfo {author} {\bibfnamefont {S.}~\bibnamefont {K.~Tripathy}},\ }\href {https://dx.doi.org/10.1209/0295-5075/ac535b} {\bibfield  {journal} {\bibinfo  {journal} {Eur. Phys. Lett.}\ }\textbf {\bibinfo {volume} {136}},\ \bibinfo {pages} {62001} (\bibinfo {year} {2022})}\BibitemShut {NoStop}%
\bibitem [{\citenamefont {Andersson}\ \emph {et~al.}(1983)\citenamefont {Andersson} \emph {et~al.}}]{andersson1983}%
  \BibitemOpen
  \bibfield  {author} {\bibinfo {author} {\bibfnamefont {B.}~\bibnamefont {Andersson}} \emph {et~al.},\ }\href {https://doi.org/10.1016/0370-1573(83)90080-7} {\bibfield  {journal} {\bibinfo  {journal} {Phys. Rept.}\ }\textbf {\bibinfo {volume} {97}},\ \bibinfo {pages} {31} (\bibinfo {year} {1983})}\BibitemShut {NoStop}%
\bibitem [{\citenamefont {Sj{\"o}strand}(1984)}]{sjostrand1984}%
  \BibitemOpen
  \bibfield  {author} {\bibinfo {author} {\bibfnamefont {T.}~\bibnamefont {Sj{\"o}strand}},\ }\href {https://doi.org/10.1016/0550-3213(84)90607-2} {\bibfield  {journal} {\bibinfo  {journal} {Nucl. Phys. B}\ }\textbf {\bibinfo {volume} {248}},\ \bibinfo {pages} {469} (\bibinfo {year} {1984})}\BibitemShut {NoStop}%
\bibitem [{\citenamefont {Biro}\ \emph {et~al.}(1984)\citenamefont {Biro}, \citenamefont {Nielsen},\ and\ \citenamefont {Knoll}}]{biro1984}%
  \BibitemOpen
  \bibfield  {author} {\bibinfo {author} {\bibfnamefont {T.}~\bibnamefont {Biro}}, \bibinfo {author} {\bibfnamefont {H.~B.}\ \bibnamefont {Nielsen}},\ and\ \bibinfo {author} {\bibfnamefont {J.}~\bibnamefont {Knoll}},\ }\href {https://doi.org/10.1016/0550-3213(84)90441-3} {\bibfield  {journal} {\bibinfo  {journal} {Nucl. Phys. B}\ }\textbf {\bibinfo {volume} {245}},\ \bibinfo {pages} {449} (\bibinfo {year} {1984})}\BibitemShut {NoStop}%
\bibitem [{\citenamefont {Bierlich}\ \emph {et~al.}(2015)\citenamefont {Bierlich} \emph {et~al.}}]{bierlich2015strings}%
  \BibitemOpen
  \bibfield  {author} {\bibinfo {author} {\bibfnamefont {C.}~\bibnamefont {Bierlich}} \emph {et~al.},\ }\href {https://doi.org/10.1007/JHEP03(2015)148} {\bibfield  {journal} {\bibinfo  {journal} {J. High Energy Phys.}\ }\textbf {\bibinfo {volume} {03}}\bibinfo  {number} { (2015)},\ \bibinfo {pages} {148}}\BibitemShut {NoStop}%
\bibitem [{\citenamefont {Bianchi}\ \emph {et~al.}(2016)\citenamefont {Bianchi} \emph {et~al.}}]{bianchi2016}%
  \BibitemOpen
\bibfield  {number} {  }\bibfield  {author} {\bibinfo {author} {\bibfnamefont {L.}~\bibnamefont {Bianchi}} \emph {et~al.} (\bibinfo {collaboration} {ALICE Collaboration}),\ }\href {https://doi.org/10.1016/j.nuclphysa.2016.03.005} {\bibfield  {journal} {\bibinfo  {journal} {Nucl. Phys. A}\ }\textbf {\bibinfo {volume} {956}},\ \bibinfo {pages} {777} (\bibinfo {year} {2016})}\BibitemShut {NoStop}%
\bibitem [{\citenamefont {Adam}\ \emph {et~al.}(2017{\natexlab{a}})\citenamefont {Adam} \emph {et~al.}}]{alice2017enhanced}%
  \BibitemOpen
  \bibfield  {author} {\bibinfo {author} {\bibfnamefont {J.}~\bibnamefont {Adam}} \emph {et~al.} (\bibinfo {collaboration} {ALICE Collaboration}),\ }\href {https://doi.org/10.1038/nphys4111} {\bibfield  {journal} {\bibinfo  {journal} {Nature Phys.}\ }\textbf {\bibinfo {volume} {13}},\ \bibinfo {pages} {535} (\bibinfo {year} {2017}{\natexlab{a}})}\BibitemShut {NoStop}%
\bibitem [{\citenamefont {Nayak}\ \emph {et~al.}(2019)\citenamefont {Nayak} \emph {et~al.}}]{nayak2019}%
  \BibitemOpen
  \bibfield  {author} {\bibinfo {author} {\bibfnamefont {R.}~\bibnamefont {Nayak}} \emph {et~al.},\ }\href {https://doi.org/10.1103/PhysRevD.100.074023} {\bibfield  {journal} {\bibinfo  {journal} {Phys. Rev. D}\ }\textbf {\bibinfo {volume} {100}},\ \bibinfo {pages} {074023} (\bibinfo {year} {2019})}\BibitemShut {NoStop}%
\bibitem [{\citenamefont {Hushnud}\ and\ \citenamefont {Dey}(2024)}]{hushnud2024}%
  \BibitemOpen
  \bibfield  {author} {\bibinfo {author} {\bibfnamefont {H.}~\bibnamefont {Hushnud}}\ and\ \bibinfo {author} {\bibfnamefont {K.}~\bibnamefont {Dey}},\ }\href {https://doi.org/10.1016/j.nuclphysa.2024.122868} {\bibfield  {journal} {\bibinfo  {journal} {Nucl. Phys. A}\ }\textbf {\bibinfo {volume} {1046}},\ \bibinfo {pages} {122868} (\bibinfo {year} {2024})}\BibitemShut {NoStop}%
\bibitem [{\citenamefont {Acharya}\ \emph {et~al.}(2024)\citenamefont {Acharya} \emph {et~al.}}]{acharya2024}%
  \BibitemOpen
  \bibfield  {author} {\bibinfo {author} {\bibfnamefont {S.}~\bibnamefont {Acharya}} \emph {et~al.} (\bibinfo {collaboration} {ALICE Collaboration}),\ }\href {https://doi.org/10.1007/JHEP09(2024)204} {\bibfield  {journal} {\bibinfo  {journal} {J. High Energy Phys.}\ }\textbf {\bibinfo {volume} {09}}\bibinfo  {number} { (2024)},\ \bibinfo {pages} {204}}\BibitemShut {NoStop}%
\bibitem [{\citenamefont {Acharya}\ \emph {et~al.}(2025)\citenamefont {Acharya} \emph {et~al.}}]{acharya2025}%
  \BibitemOpen
\bibfield  {number} {  }\bibfield  {author} {\bibinfo {author} {\bibfnamefont {S.}~\bibnamefont {Acharya}} \emph {et~al.} (\bibinfo {collaboration} {ALICE Collaboration}),\ }\href {https://doi.org/10.1103/PhysRevLett.134.022303} {\bibfield  {journal} {\bibinfo  {journal} {Phys. Rev. Lett.}\ }\textbf {\bibinfo {volume} {134}},\ \bibinfo {pages} {022303} (\bibinfo {year} {2025})}\BibitemShut {NoStop}%
\bibitem [{\citenamefont {Rustamov}\ \emph {et~al.}(2023)\citenamefont {Rustamov} \emph {et~al.}}]{rustamov2017}%
  \BibitemOpen
  \bibfield  {author} {\bibinfo {author} {\bibfnamefont {A.}~\bibnamefont {Rustamov}} \emph {et~al.},\ }\href {https://www.sciencedirect.com/science/article/pii/S0375947423000441} {\bibfield  {journal} {\bibinfo  {journal} {Nuclear Physics A}\ }\textbf {\bibinfo {volume} {1034}},\ \bibinfo {pages} {122641} (\bibinfo {year} {2023})}\BibitemShut {NoStop}%
\bibitem [{\citenamefont {Adam}\ \emph {et~al.}(2017{\natexlab{b}})\citenamefont {Adam} \emph {et~al.}}]{adam2017}%
  \BibitemOpen
  \bibfield  {author} {\bibinfo {author} {\bibfnamefont {J.}~\bibnamefont {Adam}} \emph {et~al.} (\bibinfo {collaboration} {ALICE Collaboration}),\ }\href {https://doi.org/10.1140/epjc/s10052-016-4571-1} {\bibfield  {journal} {\bibinfo  {journal} {Eur. Phys. J. C}\ }\textbf {\bibinfo {volume} {77}},\ \bibinfo {pages} {33} (\bibinfo {year} {2017}{\natexlab{b}})}\BibitemShut {NoStop}%
\bibitem [{\citenamefont {Ortiz}(2021)}]{ortiz2021}%
  \BibitemOpen
  \bibfield  {author} {\bibinfo {author} {\bibfnamefont {A.}~\bibnamefont {Ortiz}},\ }\href {https://link.aps.org/doi/10.1103/PhysRevD.104.076019} {\bibfield  {journal} {\bibinfo  {journal} {Phys. Rev. D}\ }\textbf {\bibinfo {volume} {104}},\ \bibinfo {pages} {076019} (\bibinfo {year} {2021})}\BibitemShut {NoStop}%
\bibitem [{\citenamefont {Bencédi}\ \emph {et~al.}(2020)\citenamefont {Bencédi}, \citenamefont {Ortiz},\ and\ \citenamefont {Tripathy}}]{bencedi2021}%
  \BibitemOpen
  \bibfield  {author} {\bibinfo {author} {\bibfnamefont {G.}~\bibnamefont {Bencédi}}, \bibinfo {author} {\bibfnamefont {A.}~\bibnamefont {Ortiz}},\ and\ \bibinfo {author} {\bibfnamefont {S.}~\bibnamefont {Tripathy}},\ }\href {https://dx.doi.org/10.1088/1361-6471/abc5fb} {\bibfield  {journal} {\bibinfo  {journal} {J. Phys. G: Nucl. Part. Phys.}\ }\textbf {\bibinfo {volume} {48}},\ \bibinfo {pages} {015007} (\bibinfo {year} {2020})}\BibitemShut {NoStop}%
\bibitem [{\citenamefont {Tumasyan}\ \emph {et~al.}(2023)\citenamefont {Tumasyan} \emph {et~al.}}]{tumasyan2023cms}%
  \BibitemOpen
  \bibfield  {author} {\bibinfo {author} {\bibfnamefont {A.}~\bibnamefont {Tumasyan}} \emph {et~al.} (\bibinfo {collaboration} {CMS Collaboration}),\ }\href {https://doi.org/10.1140/epjc/s10052-023-11630-8} {\bibfield  {journal} {\bibinfo  {journal} {Eur. Phys. J. C}\ }\textbf {\bibinfo {volume} {83}},\ \bibinfo {pages} {587} (\bibinfo {year} {2023})}\BibitemShut {NoStop}%
\bibitem [{\citenamefont {Thakur}\ \emph {et~al.}(2023)\citenamefont {Thakur} \emph {et~al.}}]{thakur2023}%
  \BibitemOpen
  \bibfield  {author} {\bibinfo {author} {\bibfnamefont {J.}~\bibnamefont {Thakur}} \emph {et~al.},\ }\href {https://doi.org/10.1016/j.nuclphysa.2023.122659} {\bibfield  {journal} {\bibinfo  {journal} {Nucl. Phys. A}\ }\textbf {\bibinfo {volume} {1035}},\ \bibinfo {pages} {122659} (\bibinfo {year} {2023})}\BibitemShut {NoStop}%
\bibitem [{\citenamefont {Vechernin}(2015)}]{vechernin2015forward}%
  \BibitemOpen
  \bibfield  {author} {\bibinfo {author} {\bibfnamefont {V.}~\bibnamefont {Vechernin}},\ }\href {https://doi.org/10.1016/j.nuclphysa.2015.03.009} {\bibfield  {journal} {\bibinfo  {journal} {Nucl. Phys. A}\ }\textbf {\bibinfo {volume} {939}},\ \bibinfo {pages} {21} (\bibinfo {year} {2015})}\BibitemShut {NoStop}%
\bibitem [{\citenamefont {Engel}\ and\ \citenamefont {Ranft}(1996)}]{engel1996}%
  \BibitemOpen
  \bibfield  {author} {\bibinfo {author} {\bibfnamefont {R.}~\bibnamefont {Engel}}\ and\ \bibinfo {author} {\bibfnamefont {J.}~\bibnamefont {Ranft}},\ }\href {https://doi.org/10.1103/PhysRevD.54.4244} {\bibfield  {journal} {\bibinfo  {journal} {Phys. Rev. D}\ }\textbf {\bibinfo {volume} {54}},\ \bibinfo {pages} {4244} (\bibinfo {year} {1996})}\BibitemShut {NoStop}%
\bibitem [{\citenamefont {Skands}(2010)}]{skands2010}%
  \BibitemOpen
  \bibfield  {author} {\bibinfo {author} {\bibfnamefont {P.}~\bibnamefont {Skands}},\ }\href {https://doi.org/10.1103/PhysRevD.82.074018} {\bibfield  {journal} {\bibinfo  {journal} {Phys. Rev. D}\ }\textbf {\bibinfo {volume} {82}},\ \bibinfo {pages} {074018} (\bibinfo {year} {2010})}\BibitemShut {NoStop}%
\bibitem [{\citenamefont {Aad}\ \emph {et~al.}(2011)\citenamefont {Aad} \emph {et~al.}}]{aad2011}%
  \BibitemOpen
  \bibfield  {author} {\bibinfo {author} {\bibfnamefont {G.}~\bibnamefont {Aad}} \emph {et~al.} (\bibinfo {collaboration} {ATLAS Collaboration}),\ }\href {https://dx.doi.org/10.1088/1367-2630/13/5/053033} {\bibfield  {journal} {\bibinfo  {journal} {New J. Phys.}\ }\textbf {\bibinfo {volume} {13}},\ \bibinfo {pages} {053033} (\bibinfo {year} {2011})}\BibitemShut {NoStop}%
\bibitem [{\citenamefont {Khachatryan}\ \emph {et~al.}(2017)\citenamefont {Khachatryan} \emph {et~al.}}]{khachatryan2017}%
  \BibitemOpen
  \bibfield  {author} {\bibinfo {author} {\bibfnamefont {V.}~\bibnamefont {Khachatryan}} \emph {et~al.} (\bibinfo {collaboration} {CMS Collaboration}),\ }\href {https://doi.org/10.1016/j.physletb.2016.12.009} {\bibfield  {journal} {\bibinfo  {journal} {Phys. Lett. B}\ }\textbf {\bibinfo {volume} {765}},\ \bibinfo {pages} {193} (\bibinfo {year} {2017})}\BibitemShut {NoStop}%
\end{thebibliography}

\providecommand{\noopsort}[1]{}\providecommand{\singleletter}[1]{#1}%
%



\end{document}